\documentclass[amstex,aps,showpacs,pra,preprint,nofootinbib]{revtex4}%
\usepackage{graphicx}
\usepackage{color}
\usepackage{bm}
\usepackage{longtable}
\usepackage{amsmath}
\usepackage{amsfonts}
\usepackage{amssymb}

\begin{document}
\title{Weak measurements of trajectories in quantum systems: classical, Bohmian and
sum over paths}
\author{A. Matzkin}
\affiliation{Laboratoire de Physique Th\'{e}orique et Mod\'{e}lisation (CNRS Unit\'{e}
8089), Universit\'{e} de Cergy-Pontoise, 95302 Cergy-Pontoise cedex, France}

\begin{abstract}
Weak values, obtained from weak measurements, attempt to describe the
properties of a quantum system as it evolves from an initial to a final state,
without practically altering this evolution. Trajectories can be defined from
weak measurements of the position, or inferred from weak values of the
momentum operator. The former can be connected to the asymptotic form of the
Feynman propagator and the latter to Bohmian trajectories. Employing a
time-dependent oscillator as a model, this work analyzes to what extent weak
measurements can shed light on the underlying dynamics of a quantum system
expressed in terms of trajectories, in particular by comparing the two approaches.

\end{abstract}

\pacs{03.65.Ta, 03.65.Ca}
\maketitle

\section{Introduction}

Contrary to classical physics, the standard formalism of quantum mechanics
forbids the use of space-time trajectories to describe the time evolution of a
system. However trajectories surreptitiously sneak back into the description,
the interpretation and the computation of quantum phenomena. There are various
form of trajectories that have been found useful.\ Among these the most
prominent are the paths of the path integral approach due to Feynman
\cite{schulman} and the trajectories built on the probability flow employed in
the de Broglie-Bohm model \cite{bohm}. Both types of trajectories have been
employed to interpret experimental results.

The path integral approach has been extremely successful for quantum systems
in the semiclassical regime.\ Indeed, in this regime the path integral becomes
essentially a coherent sum over the classical trajectories of the
corresponding classical system \cite{schulman}. Such classical trajectories
have been employed to understand the properties of these systems in the
framework of ``quantum chaos'' \cite{QCbook}, and their manifestations have
have been experimentally observed in many quantum systems (see eg Refs
\cite{chaos}). The trajectories of the Bohmian model are essentially obtained
by following the probability current density arising from the Schr\"{o}dinger
equation. Bohmian trajectories have also been employed to interpret the
dynamics in several systems \cite{dbbapp}. They have further been used as a
numerical computation method, especially in molecular physics or when mixing
classical and quantum degrees of freedom in a mean field approximation is
required \cite{wyatt}.

The trajectories of the path integral, generated by the classical
Lagrangian, are generically different from the quantum trajectories
built on the Schr\"{o}dinger probability flow \cite{mismatch}. This
is not a problem as long as one sees these trajectories as being
computational tools or mathematical artefacts. However recently the
approach of weak measurements, introduced some time ago by Aharonov,
Albert and Vaidman \cite{AAV} has been employed to measure
non-perturbatively trajectories in quantum systems. The main idea
underlying weak measurements \cite{WMref1,WMref2,WMref3} is to
access the properties of a quantum system evolving from a given
initial state towards a final state, practically without disturbing
the system evolution.\ This cannot be achieved by a standard
projective measurement (doing so would irremediably disturb the
system by projective its premeasurement state to a subspace spanned
by the eigenstate of the measured observable). Instead, the system
observable is coupled unitarily to an ancilla acting as a weak
measurement apparatus (WMA). If the coupling is weak, the unitary
evolution of the system is practically left unperturbed. The value
recorded by a WMA is not an eigenvalue (since there is no state
projection) but what is known as a \emph{weak value} \cite{AAV} of
the weakly measured observable.

The idea of inferring Bohmian trajectories from weak measurements of the
\emph{momentum} (followed by a projective measurement of the position) has
been proposed first by Leavens \cite{leavens} (see also
\cite{wiseman07,hiley,robust} for more recent approaches). This
scheme was experimentally implemented in a two-slit interferometer
\cite{kocsis} allowing to reconstruct Bohmian trajectories from the observed
data.\ A method that can in principle allow to observe the Feynman paths with
weak measurements of the \emph{position} (including the coherent paths
superposition) was also suggested recently \cite{weak trajectories} (see also
\cite{tsutsui}). In a
first view it is therefore tempting to conclude that the type of trajectory
that one sees depends eventually on what is being measured, which in turn
calls for a definite experimental setup.

The aim of this work is to examine this question in details by displaying
expressions for the weak measurement of classical and Bohmian trajectories in
the same system. We will employ a tractable model system -- a two dimensional
time-dependent linear oscillator (TDLO). While the dynamics of the TDLO is
arguably less rich than that of generic systems, its main advantage in the
present context is that the Feynman sum over paths can be obtained exactly in
closed form (without invoking the semiclassical approximation) while the
computation of the Bohmian trajectories is numerically tractable. At the same
time the time-dependent aspect allows to "simulate" dynamical feature that
generally appear in systems with more involved dynamics (like recurrences of
closed orbits). Moreover the TDLO has often been employed to model quantum
systems such as the dynamics of trapped ions \cite{rmp}, photon generation in
quantum optics \cite{photon} or cosmological mini-superspace models
\cite{cosmo}.

The paper is organized as follows.\ We first introduce the weak measurements
framework and derive the two types of trajectories that can be inferred from
weak measurements. We then introduce our model system and detail how the
quantum dynamics can be interpreted in terms of classical or Bohmian
trajectories (Sec.\ III), yielding different interpretations of dynamical
phenomena when classical and Bohmian trajectories differ. Sec.\ IV describes
the trajectories inferred from weak measurements for the TDLO, including
derivations and several numerical illustrations for specific cases. We discuss
our findings and conclude in Sec.\ V, while an Appendix details how we obtain
the closed form solutions of the Schr\"{o}dinger equation for the TDLO.

\section{Weak Measurements}

\subsection{Weak measurement framework}

The underlying idea at the basis of the weak measurement framework
is an attempt to answer the question:"\emph{what is the value of a
property (represented by an observable $\hat{A}$) of a quantum
system while it is evolving from an initial state $\left\vert
\psi(t_{0})\right\rangle $ to a final state $\left\vert
\chi\right\rangle $?}". Instead of making a projective measurement
the observable $\hat{A}$ is coupled unitarily to a dynamical
variable of an ancilla (the weak measurement apparatus, WMA).\ There
is thus no projection of the system's quantum state at this stage.
Moreover if the coupling is asymptotically weak, it can be shown
that the state of the system is left practically undisturbed. The
system thus continues\footnote{Formally, the system and the WMA get
entangled, so the coupling interaction changes the dynamics of the
overall entangled wavefunction. For a small coupling, the
postselection probabilities are not modified to first order while
the WMA's wavefunction picks up a shift; see eg Secs 2 and 5 of
\cite{WMref1}, and \cite{pan12} for a problem solved with exact
solutions of the Schr\"{o}dinger equations.} its evolution until a
final projective measurement (of another observable $\hat{B}$)
projects its state to $\left\vert \chi\right\rangle $, the
post-selected state. As a result of the unitary coupling, a
projective measurement on the system also modifies the quantum state
of the WMA: the variable conjugate to the coupled one is shifted by
a quantity proportional to $\operatorname{Re}\left\langle
A_{w}\right\rangle$ where $\left\langle A _{w}\right\rangle$ is the
\emph{weak value }of the observable $\hat{A}$ when the system is
pre-selected in state $\left\vert \psi(t_{0})\right\rangle $ and
post-selected to the state $\left\vert \chi\right\rangle .$ Letting
$U(t^{\prime\prime},t^{\prime})$ denote the evolution operator of
the system between times $t^{\prime}$ and
$t^{\prime\prime}$, $\left\langle \hat{A}_{w}\right\rangle $ is given by%
\begin{equation}
\left\langle \hat{A}_{w}\right\rangle =\frac{\left\langle \chi\right\vert
U(t_{f},t_{w})\hat{A}\left[  U(t_{w},t_{0})\left\vert \psi(t_{0})\right\rangle
\right]  }{\left\langle \chi\right\vert U(t_{f},t_{0})\left\vert \psi
(t_{0})\right\rangle },\label{wv-def}%
\end{equation}
where $t_{w}$ and $t_{f}$ stand for the times at which the weak
measurement and final postselection take place respectively.\
Introducing the notation $\left\langle \chi(t_{w})\right\vert%
\equiv\left\langle \chi\right\vert U (t_{f},t_{w})$ representing the
postselected state evolved backward in time, while $\left\vert
\psi(t_{w})\right\rangle =U(t_{w},t_{0})\left\vert
\psi(t_{0})\right\rangle ,$ the weak value (\ref{wv-def}) is written
in terms
of quantities taken at the weak measurement time $t_{w}$ as%
\begin{equation}
\left\langle \hat{A}_{w}\right\rangle =\frac{\left\langle \chi(t_{w}%
)\right\vert \hat{A}\left\vert \psi(t_{w})\right\rangle }{\left\langle
\chi(t_{w})\right\vert \left.  \psi(t_{w})\right\rangle }.\label{wvt}%
\end{equation}

Hence the WMA acts as a pointer that records the weak value
\footnote{The pointer can actually register either the real or the
imaginary parts of the weak value, which is a complex number. Only
the real part is related to the value of the measured observable,
while the imaginary part is typically related to the measurement
backaction \cite{joza}.} of the weakly measured observable. While
there has been a controversy on the meaning of weak values from
their inception, in our view the controversy has more to do with the
interpretation given to the theoretical terms of quantum theory in
general than to the specificities of weak measurements as such.
There is now ample evidence \cite{WMref1,WMref2} that the weak
values given by Eq. (\ref{wv-def}) capture a universal effect in
which $\operatorname{Re}\left\langle A_{w}\right\rangle $ represents
the response on the probe of a minimally disturbing interaction for
the system reflecting the value of the property described by
$\hat{A}$ relative to the fraction of the initial state that will
conditionally end up in the post-selected state $\left\vert
\chi\right\rangle $. The basic property allowing this interpretation
follows by writing the expectation value of $\hat{A}$ \ when the
system is in the state $\left\vert \psi(t_{w})\right\rangle $ in
terms of the probabilities of reaching each eigenstate $\left\vert
\chi_{f}\right\rangle $ of a different observable $\hat{B},$
\begin{equation}
\left\langle \psi(t_{w})\right\vert \hat{A}\left\vert \psi(t_{w})\right\rangle
=\sum_{f}\left\vert \left\langle \chi_{f}\right\vert \left.  \psi
(t_{w})\right\rangle \right\vert ^{2}\operatorname{Re}\frac{\left\langle
\chi_{f}\right\vert \hat{A}\left\vert \psi(t_{w})\right\rangle }{\left\langle
\chi_{f}\right\vert \left.  \psi(t_{w})\right\rangle },\label{wv prob}%
\end{equation}
where the restriction to the real part comes from the fact that
since the left-handside is real, the sum over the imaginary parts
vanishes. The interpretation of this formula in terms of weak
measurements assumes impilictly that the probabilities of obtaining
the final state $\left\vert \chi_{f}\right\rangle $ are not modified
by the action of $\hat{A}$.

Note that in terms of projective measurements, Eq. (\ref{wv prob}) can also be
read similarly to the standard quantum mechanical expectation value expression%
\begin{equation}
\left\langle \psi(t_{w})\right\vert \hat{A}\left\vert \psi(t_{w})\right\rangle
=\sum_{f}\left\vert \left\langle \alpha_{f}\right\vert \left.  \psi
(t_{w})\right\rangle \right\vert ^{2}\alpha_{f}\label{ev}%
\end{equation}
where $\hat{A}\left\vert \alpha_{f}\right\rangle =\alpha_{f}\left\vert
\alpha_{f}\right\rangle .\ $While Eq. (\ref{ev}) involves a protocol in which
the expectation value is obtained by measuring the eigenvalues $\alpha_{f}$ of
$\hat{A}$ and their relative frequencies, Eq. (\ref{wv prob}) suggests a
protocol in which the expectation value of $\hat{A}$ is obtained by a weak
measurement of $\hat{A}$ followed by a standard projective measurement of
$\hat{B},$ for which only the relative frequencies for obtaining the
eigenvalues $\chi_{f}$ of $\hat{B}$ are needed.

\subsection{Weak measurements of position: "weak" trajectories}

The most intuitive way of measuring a trajectory follows from its definition:
in a given frame of reference the position $\mathbf{r}(t)$ is recorded as a
function of time. For an evolving quantum system, this involves monitoring the
position not only non-destructively, but without affecting the subsequent
evolution of the system. A weak measurement of the position is perfectly
suited in order to monitor the position. Starting from a localized initial
state $\left\vert \psi(t_{0})\right\rangle $ -- this will be the preselected
state --, and ending with a projective measurement to a state localized at a
given final position at time $t_{f}$ (this will be the postselected state), we
can place a series of weak measurement apparata (WMA) that weakly interact
with the system via a local coupling to the position observable $\mathbf{\hat
{r}}$.

Assume first that there is a single WMA, lying at position $\mathbf{R}^{0}$
and whose wavefunction $\phi(\mathbf{R)}$ is tightly localized around the
central position $\mathbf{R}_{{}}^{0}$, eg a Gaussian wavefunction
$\phi(\mathbf{R}_{{}})=(2/\pi\Delta^{2})^{1/2}e^{-\left(  \mathbf{R}%
-\mathbf{R}^{0}\right)  ^{2}/\Delta^{2}}$. The weak interaction is triggered
when the system wavefunction enters the region around $\mathbf{R}^{0}$,
corresponding to a contact interaction of the form
\begin{equation}
H^{0}=\gamma(t)\mathbf{\hat{r}}\cdot\mathbf{R}^{0}f\left(  \left\vert
\mathbf{\hat{r}}-\mathbf{R}^{0}\right\vert ^{2}\right) \label{int}%
\end{equation}
where $\gamma$ is a smooth function of $t$ determining the coupling and $f$ is
a function sharply peaked at $\left\vert \mathbf{\hat{r}}-\mathbf{R}%
^{0}\right\vert =0$ indicating the short-range character of the contact
interaction between the system and the WMA. Let $t_{w}$ denote the mean time
at which the interaction takes place. If the duration $\tau$ of the
measurement is short relative to the timescale of the system dynamics, then
\cite{supplem} the WMA records \footnote{Since the spatial WMA wavefunction
picks up an $\mathbf{R}$ dependent phase term proportional to the weak value,
the WMA pointer is monitored in momentum space.} the weak value of the
position at time $t_{w}$
\begin{equation}
\left\langle \mathbf{r}(t_{w})\right\rangle _{W}\equiv\frac{\left\langle
\chi(t_{w})\right\vert \mathbf{\hat{r}}f\left(  \left\vert \mathbf{\hat{r}%
}-\mathbf{R}^{0}\right\vert ^{2}\right)  \left\vert \psi(t_{w})\right\rangle
}{\left\langle \chi(t_{w})\right\vert \left.  \psi(t_{w})\right\rangle
}.\label{wp}%
\end{equation}
We will see below that $\left\langle \mathbf{r}(t_{w})\right\rangle _{W}$ can
take a simple form for specific choices of the system wavefunction.
Nevertheless one can see qualitatively that the wavefunctions $\psi
(t_{w},\mathbf{r})$ and $\chi(t_{w},\mathbf{r})$ must overlap significantly
around the position of the WMA $\mathbf{r\approx R}^{0}$ in order to obtain a
non vanishing weak value. If $\left\langle \mathbf{r}(t_{w})\right\rangle
_{W}=0, $ this essentially means there is no wavefunction exploring the region
around $\mathbf{R}^{0}$ compatible with the postselected state. In the special
case in which $\psi(t_{w},\mathbf{r})$ (the preselected state propagated
forward in time) and $\chi(t_{w},\mathbf{r})$ (the postselected state
propagated backward in time) are identical, then it is easy to see that under
certain conditions (eg, $\psi$ and $\chi$ are constant in the region where
$f\left(  \left\vert \mathbf{\hat{r}}-\mathbf{R}^{0}\right\vert ^{2}\right)  $
is non-zero, or have a maximum at $\mathbf{R}^{0}$) the weak value will be
simply given by $\left\langle \mathbf{r}(t_{w})\right\rangle _{W}%
=\mathbf{R}^{0}$, that is the location of the WMA.

Assume now there are several WMA of the same type distributed at positions
$\mathbf{R}_{k}^{0},$ $k=1,...,N$. It is convenient to label them according to
the order in which they interact ($k=1$ corresponds to the meter interacting
first with the system, $k=2$ to the second meter having interacted with the
system and so on). Each WMA, endowed with its own wavefunction $\phi
(\mathbf{R}_{k}\mathbf{)}$ localized around $\mathbf{R}_{k}^{0}$ interacts at
time $t_{k}$ with the wavefunction through a contact interaction of the form
(\ref{int}) and registers a weak value%
\begin{equation}
\left\langle \mathbf{r}(t_{k});\mathbf{R}_{k}^{0}\right\rangle _{W}\equiv
\frac{\left\langle \chi(t_{k})\right\vert \mathbf{\hat{r}}f\left(  \left\vert
\mathbf{\hat{r}}-\mathbf{R}_{k}^{0}\right\vert ^{2}\right)  \left\vert
\psi(t_{k})\right\rangle }{\left\langle \chi(t_{k})\right\vert \left.
\psi(t_{k})\right\rangle }\label{wvwt}%
\end{equation}
where $\mathbf{R}_{k}^{0},$ labeling the position of the meter, will be
omitted in most of the text. Overall, out of the $N$ WMA that act as meters
recording the weak values, only $n$ will display a non-zero value, those for
which postselection is compatible with the dynamics of the preselected state
at the given WMA positions.\ Relabeling $k$ in a time-ordered manner
reflecting the times at which the WMA have interacted, the set
\begin{equation}
\text{WT}_{\psi(t_{0}),\chi(t_{f})}=\{t_{k},\operatorname{Re}\left[
\left\langle \mathbf{r}(t_{k})\right\rangle _{W}\right]  \}, \; \;
k=1,...n\label{wt}%
\end{equation}
\ defines a trajectory in the sense of weak position measurements, that is a
``weak trajectory'' for given pre and postselected states. Note that the pre
and post-selected states $\left\vert \psi(t_{k})\right\rangle $ and
$\left\vert \chi(t_{k})\right\rangle $ at times $t_{k}$ cannot be freely
chosen but depend on the initial pre-selected state $\left\vert \psi
(t_{0})\right\rangle $ and on the final post-selected state $\left\vert
\chi(t_{f})\right\rangle $ respectively.

For an arbitrary quantum system a WT (\ref{wt}) will typically reflect the
space-time correlation between the forward evolution of the preselected state
and the backward evolution of the postselected state at the positions
$\mathbf{R}_{k}^{0}$ of the weakly interacting meters: only the WMA at
positions for which this correlation is non-vanishing will display a non-zero
weak value. WTs become particularly interesting in the semiclassical regime,
ie when the Feynman path integral is approximately given by the semiclassical
propagator involving a propagator given by a coherent sum over the paths of
the classical corresponding system. Indeed, as we will see below (Secs.\ III
and IV) the weak trajectories can in principle be employed to record the sum
over paths of the semiclassical propagator. But we will first introduce
another type of trajectories that can be inferred from a different type of
weak measurement.

\subsection{Weak measurements of momentum: velocity field\label{wmbt}}

The standard textbook form of the quantum mechanical probability current for a
system in state $\psi(\mathbf{r},t)$ is given by%
\begin{equation}
\mathbf{j}(\mathbf{r},t)=\frac{i\hslash}{2m}\left[  \psi(\mathbf{r}%
,t)\nabla\psi^{\ast}(\mathbf{r},t)-\psi^{\ast}(\mathbf{r},t)\nabla
\psi(\mathbf{r},t)\right]  .\label{cur}%
\end{equation}
A local velocity field at the space-time point $(\mathbf{r},t)$ can be defined
from the current density through
\begin{equation}
\mathbf{v}(\mathbf{r},t)=\frac{\mathbf{j}(\mathbf{r},t)}{\rho(\mathbf{r}%
,t)}\label{vel}%
\end{equation}
where $\rho(\mathbf{r},t)\equiv\left\vert \psi(\mathbf{r},t)\right\vert ^{2}$.
Consider now applying the weak value definition (\ref{wvt}) to a weak
measurement of the momentum operator $\mathbf{\hat{p}}$ when the system is in
state $\left\vert \psi(t)\right\rangle $ immediately followed by a
postselection to the position eigenstate $\left\vert \mathbf{r}_{f}%
\right\rangle $. The real and imaginary parts of the weak value are given by
\begin{align}
\left\langle \mathbf{p}(t)\right\rangle _{W}  &  =\frac{\left\langle
\mathbf{r}_{f}\right\vert \mathbf{\hat{p}}\left\vert \psi(t)\right\rangle
}{\left\langle \mathbf{r}_{f}\right\vert \left.  \psi(t)\right\rangle
}\label{wm-mom}\\
&  =m\mathbf{v}(\mathbf{r}_{f},t)-i\hslash\frac{\mathbf{\triangledown}%
\rho(\mathbf{r}_{f},t)}{2\rho(\mathbf{r}_{f},t)}.\label{wm-vel}%
\end{align}
Hence by performing weak measurements of the momentum at different space-time
points $(\mathbf{r}_{f},t)$, the real part of the weak value $\left\langle
\mathbf{p}(t)\right\rangle _{W}$ allows to reconstruct the velocity field
$\mathbf{v}(\mathbf{r},t)$.

Note that  $\left\langle \mathbf{p}(t)\right\rangle _{W}$ can also
be obtained from the difference of two position measurements made in
a very small time interval \cite{wiseman07}. One first defines a
weak position measurement at time $t-\varepsilon$ consistent with
postselection at position  $\mathbf{r}_{f}$ at time $t$:
\begin{equation}
\left\langle \mathbf{r}(t-\varepsilon)\right\rangle _{W}=\frac{\left\langle \mathbf{r}%
_{f}\right\vert U(t,t-\varepsilon)\mathbf{\hat{r}}\left\vert
\psi(t-\varepsilon)\right\rangle }{\left\langle
\mathbf{r}_{f}\right\vert U(t,t-\varepsilon)\left\vert
\psi(t-\varepsilon)\right\rangle }.
\end{equation}
To first order in $\varepsilon\rightarrow0$ one obtains after some
manipulations \cite{wiseman07}
\begin{equation}
\left\langle \mathbf{p}(t)\right\rangle _{W}=m\frac{\left(
\mathbf{r}_{f}(t)-\left\langle
\mathbf{r}(t-\varepsilon)\right\rangle _{W}\right)}{\varepsilon}
\label{av}%
\end{equation}
so that the velocity field appears as the real part of the
difference $\left( \mathbf{r}_{f}(t)-\left\langle
\mathbf{r(t-\varepsilon)}\right\rangle _{W}\right)  /\varepsilon $.

It turns out \cite{leavens} that this velocity field matches the local
particle velocity\ in the de Broglie-Bohm interpretation of quantum mechanics
(see Sec.\ \ref{sbohmian} below). The imaginary part of the weak value
(\ref{wm-vel}) has also very recently \cite{hiley} been interpreted as the
osmotic velocity of a putative stochastic model underlying the de Broglie-Bohm
framework. The relations (\ref{wm-mom})-(\ref{wm-vel}) between weak values and
the Bohmian momentum are important, not only because they allow in principle
to extract experimentally Bohmian trajectories from the weak measurements of
the momentum velocity field, but also because these relations constitute a
link between the momentum operator and the Bohmian particles momentum (the
latter having little to do with the eigenvalues of the momentum operator).

\section{Quantum properties and trajectories}

\subsection{Model system and setting}

As mentioned in the Introduction, it is often useful to interpret properties
of quantum systems in terms of trajectories. This is particularly the case for
quantum systems in the semiclassical regime for which the underlying classical
dynamics drives the quantum evolution operator. Nevertheless a typical system
in the semiclassical regime is not easy to handle -- the full semiclassical
propagation is generally a formidable task, the search for classical periodic
orbits in general is not trivial and the computation of Bohmian trajectories
calls for a powerful numerical implementation. We will work instead with a
simple model system, a two dimensional time-dependent linear oscillator
(TDLO).\ The TDLO allows to simulate the sum over paths aspect of the
semiclassical propagator, given below by Eq. (\ref{e10}) in an easy and
tractable manner. Moreover since the TDLO\ Hamiltonian (\ref{ham2D}) is
quadratic, the semiclassical approximation is quantum mechanically exact
\cite{schulman}.\

We shall consider in the following a two-dimensional time-dependent oscillator
with time-dependent frequencies $V_{x}(t)$ and $V_{y}(t)$, whose Hamiltonian
is given by
\begin{equation}
H(t)=\frac{\mathbf{p}^{2}}{2m}+\frac{m}{2}\left(  V_{x}(t)x^{2}+V_{y}%
(t)y^{2}\right)  ,\label{ham2D}%
\end{equation}
where for definiteness the time-dependence of the potential will be chosen to
take the form%
\begin{equation}
V_{j}(t)=v_{j}-\kappa_{j}\cos\left(  2\omega_{j}t\right)  ,\text{ \qquad
}j=x,y\text{.}%
\end{equation}
(see Appendix A for details). Let us take an initial state made up of a single
2D\ Gaussian%
\begin{equation}
\psi^{(\mathbf{q}_{0},\mathbf{p}_{0})}(\mathbf{r},t_{0})=\left(  \frac{2m}%
{\pi\alpha_{0}^{2}}\right)  ^{1/2}e^{-m\left(  \mathbf{r}-\mathbf{q}%
_{0}\right)  ^{2}/\alpha_{0}^{2}}e^{i\mathbf{p}_{0}\cdot\left(  \mathbf{r}%
-\mathbf{q}_{0}\right)  /\hbar}\label{psi2D}%
\end{equation}
where $\mathbf{r}=(x,y),$ the parameters and $\mathbf{q}_{0}$ and
$\mathbf{p}_{0}$ are the average values of the position and momentum operators
respectively in that state; $\alpha_{0}^{{}}$ sets the width of the initial
Gaussian (for simplicity, the same initial width is taken along both
directions). Eq. (\ref{ham2D}) is a separable problem, so the solutions to the
time-dependent Schr\"{o}dinger equation are readily obtained from the 1D TDLO
described in Appendix A, yielding
\begin{equation}
\psi^{(\mathbf{q}_{0},\mathbf{p}_{0})}(\mathbf{r},t)=\psi(x,t)\psi(y,t)
\end{equation}
where each 1D wavefunction is given by Eq. (\ref{psisol}). As discussed in
Appendix A, $\psi^{(\mathbf{q}_{0},\mathbf{p}_{0})}(\mathbf{r},t)$ can be
written in terms of a wavefunction whose probability density has a maximum
along the classical trajectory $\mathbf{q}(t)=\left(  q_{x}(t),q_{y}%
(t)\right)  $ having initial position $\mathbf{q}(t)=\mathbf{q}_{0}$ and
momentum $\mathbf{p}(t)=\mathbf{p}_{0}$. The phase also depends on the
momentum $\mathbf{p}(t)$ of that classical trajectory. Two other purely
time-dependent functions $\mathbf{\alpha}(t)=(\alpha_{x}(t),\alpha_{y}(t))$
and $\mathbf{\phi}(t)=(\phi_{x}(t),\phi_{y}(t))$ linked to $\mathbf{q}(t)$ in
the framework of Ermakov systems [cf.\ Eqs (\ref{er1}) and (\ref{er2})] are
also necessary to describe the time-dependent solution, that takes the form%
\begin{align}
\psi^{(\mathbf{q}_{0},\mathbf{p}_{0})}(\mathbf{r},t)  &  =\left(  \frac
{4m}{\pi^{2}}\det[\operatorname{Re}(\mathbf{M(\alpha)})]\right)
^{1/4}\nonumber\\
&  e^{-\left[  \mathbf{r}-\mathbf{q}(t)\right]  \cdot\mathbf{M(\alpha)}%
\cdot\left[  \mathbf{r}-\mathbf{q}(t)\right]  }e^{i\mathbf{p}(t)\cdot\left[
\mathbf{x}-\mathbf{q}(t)\right]  /\hbar}e^{\frac{i}{2\hbar}\left[
\mathbf{p}(t)\cdot\mathbf{q}(t)-\mathbf{p}_{0}\cdot\mathbf{q}_{0}\right]
}e^{-\frac{i}{2}\left[  \mathbf{\phi}(t)-\mathbf{\phi}_{0}\right]
\cdot\mathbf{r}/r}.\label{psi2Dt}%
\end{align}
$\mathbf{M(\alpha)}$ is a matrix defined by%
\begin{equation}
\mathbf{M(\alpha)}=m\left(
\begin{array}
[c]{cc}%
\frac{1}{\alpha_{x}(t)^{2}}-\frac{i\alpha_{x}^{\prime}(t)}{2\hbar\alpha
_{x}(t)} & 0\\
0 & \frac{1}{\alpha_{y}(t)^{2}}-\frac{i\alpha_{y}^{\prime}(t)}{2\hbar
\alpha_{y}(t)}%
\end{array}
\right)  .
\end{equation}
Note that $\mathbf{\alpha}(t)$ determines the time-dependent width of the
evolving wavefunction; the initial state (\ref{psi2D}) corresponds to
$\mathbf{\alpha}(t_{0})=(\alpha_{0},\alpha_{0})$ and $\mathbf{\phi}%
(t_{0})=(0,0)$.

Initial states can also be superpositions of states (\ref{psi2D}); we will be
find useful to consider initial states given by%
\begin{equation}
\psi(\mathbf{r},t_{0})=\sum_{J}a_{J}\psi^{(\mathbf{q}_{0},\mathbf{p}_{0}^{J}%
)}(\mathbf{r},t_{0})\label{10}%
\end{equation}
that is a superposition (with normalized real weights $a_{J})$ of Gaussians
initially localized at the same position $\mathbf{q}_{0}$ but with different
initial mean momenta $\mathbf{p}_{0}^{J}$. The resulting wavefunction
\begin{equation}
\psi(\mathbf{r},t)=\sum_{J}a_{J}\psi^{(\mathbf{q}_{0},\mathbf{p}_{0}^{J}%
)}(\mathbf{r},t)\label{12}%
\end{equation}
is a sum of Gaussians (\ref{psi2Dt}) each propagating by following the guiding
trajectory $\mathbf{q}^{J}(t).$

\subsection{Path integral and classical trajectories}

The solutions of the Schr\"{o}dinger equation (\ref{psisol}) and
(\ref{psi2Dt}) for the TDLO in which the wavefunction amplitude is
concentrated along the trajectories of the classical corresponding system can
best be seen to arise from the Feynman path integral approach.\ From a
qualitative standpoint, the argument starts from the path integral form of the
time evolution operator
\begin{equation}
K(\mathbf{r}_{1},\mathbf{r}_{0};t_{1}-t_{0})=\int_{\mathbf{r}_{0}}%
^{\mathbf{r}_{1}}\mathcal{D}\mathbf{r}(t)\exp\frac{i}{\hbar}\left[
\int_{t_{0}}^{t_{1}}\mathcal{L}dt\right]  ,\label{prop}%
\end{equation}
the propagator, that propagates the initial wavefunction along any conceivable
path, according to
\begin{equation}
\psi(\mathbf{r},t)=\int K(\mathbf{r},\mathbf{r}_{0};t-t_{0})\psi
(\mathbf{r}_{0},t_{0})dx.\label{m1}%
\end{equation}
The propagator can be expressed in terms of classical trajectories when the
action%
\begin{equation}
R(\mathbf{r}_{1},\mathbf{r}_{0};t-t_{0})=\int_{t_{0}}^{t_{1}}\mathcal{L}dt
\end{equation}
is huge relative to $\hbar$ ($\mathcal{L}$ is the Lagrangian, given here by
the 2D extension of Eq. (\ref{lag})). In that case, the integration in Eq.
(\ref{prop}) is handled \cite{schulman} with the stationary phase
approximation, and the stationary points of the action are, by Hamilton's
principle, the classical trajectories. $K$ then takes the generic form
\footnote{This generic form actually assumes that the action has isolated
non-degenerate stationary points, which is not the case here (a change of
variables is necessary).}
\begin{equation}
K(\mathbf{r}_{1},\mathbf{r}_{0};t_{1}-t_{0})=\left(  2i\pi\hbar\right)
^{-1}\sum_{k}\left\vert \det\frac{-\partial^{2}R_{k}}{\partial\mathbf{r}%
_{1}\partial\mathbf{r}_{0}}\right\vert ^{1/2}\exp\frac{i}{\hbar}\left[
R_{k}(\mathbf{r}_{1},\mathbf{r}_{0};t_{1}-t_{0})-\mu_{k}\right]  ,\label{e10}%
\end{equation}
where $k$ runs over all the classical trajectories connecting $\mathbf{x}_{0}
$ to $\mathbf{x}_{1}$ in time $t_{1}-t_{0}$. $R_{k}$ is the classical action
and the determinant gives the classical density of paths along the $k$th
classical trajectory, and $\mu_{k}$ are additional phases related to the
number of conjugate points on the trajectory. The sole approximation made in
employing the stationary phase implies neglecting the terms beyond the second
order variations along the paths of least action.\ But for quadratic
Lagrangians -- such as (\ref{lag}) -- the third order and greater order terms
vanish, so that the semiclassical propagator (\ref{e10}) is
quantum-mechanically exact.

The propagator (\ref{e10}) accounts for the fact that the maximum of the
initial wavefunction propagates along classical trajectories. The point that
remains to be explained is the functional form of the time-dependent
wavefunction (\ref{psi2Dt}). Since the Hamiltonian (\ref{ham2D}) is separable,
it is more straightforward to deal separately with two 1D propagators. Writing
the classical action in terms of the correct dependent variables $x_{1}$ and
$x_{0}$ with the help of the Ermakov phase and amplitude functions (see
Appendix) leads after some tedious manipulations to
\begin{align}
R_{cl}(x_{1},x_{0};t_{1}-t_{0})  &  =m\left\{  \left(  x_{1}^{2}\frac
{\alpha^{\prime}(t_{1})}{\alpha(t_{1})}-x_{0}^{2}\frac{\alpha^{\prime}(t_{0}%
)}{\alpha(t_{0})}\right)  +\hslash\cot\left(  \phi(t_{1})-\phi(t_{0})\right)
\left(  \frac{x_{1}^{2}}{\alpha^{2}(t_{1})}+\frac{x_{0}^{2}}{\alpha^{2}%
(t_{0})}\right)  \right. \\
&  \left.  -2x_{1}x_{0}\hslash\left[  \frac{1}{\alpha(t_{1})\alpha(t_{0}%
)\sin\left(  \phi(t_{1})-\phi(t_{0})\right)  }\right]  \right\} \label{rcl}%
\end{align}
(see also Ref. \cite{lawande} describing a method to obtain directly the
propagator from the Ermakov system solutions). The action is quadratic in
$x_{0}$ so that with initial wavefunctions of the form (\ref{1}) Eq.
(\ref{m1}) becomes a Gaussian integral quadratic in $x_{1}$; finally the
classical solution is identified in the exponent with the help of Eq.
(\ref{ct}).

\subsection{De Broglie-Bohm trajectories \label{sbohmian}}

According to the Bohmian (or de\ Broglie-Bohm) model \cite{dbb,bohm}, a
quantum system can be seen as the combination of a point-like particle guided
by a pilot wave. The wavefunction plays the role of the pilot wave, and
through its modulus, it also gives the statistical distribution of the
particle's position, thereby recovering by construction the standard
(non-relativistic) quantum mechanical probabilities and expectation values.
From a dynamical point of view, the resulting Bohmian trajectories are the
streamlines of the usual probability current density flow derived from the
Schr\"{o}dinger equation.

If we write the wavefunction in polar form as
\begin{equation}
\psi(\mathbf{r},t)=\rho^{1/2}(\mathbf{r},t)\exp(i\sigma(\mathbf{r}%
,t)/\hbar),\label{5}%
\end{equation}
the current density, Eq. (\ref{cur}) is given as
\begin{equation}
\mathbf{j}(\mathbf{r},t)=\frac{\rho(\mathbf{r},t)\mathbf{\bigtriangledown
}\sigma(\mathbf{r},t)}{m}.\label{4}%
\end{equation}
By replacing Eq. (\ref{5}) in the Schr\"{o}dinger equation it can be seen that
$\rho$ and $\sigma$ obey the equation
\begin{equation}
\frac{\partial\sigma}{\partial t}+\frac{(\mathbf{\triangledown}\sigma)^{2}%
}{2m}+V+Q=0.\label{6}%
\end{equation}
$V$ is the usual potential (here $V=\frac{m}{2}V_{x}(t)x^{2}+\frac{m}{2}%
V_{y}(t)y^{2}$) and the term
\begin{equation}
Q(\mathbf{r},t)\equiv-\frac{\hbar^{2}}{2m}\frac{\triangledown^{2}\rho}{\rho
}\label{qp}%
\end{equation}
is known as the quantum potential. The velocity field introduced
above [Eq. (\ref{vel})] gives the velocity of the Bohmian particle
at the space-time
point $(\mathbf{r},t)$; it can be written in the form%
\begin{equation}
\mathbf{v}(\mathbf{r},t)=\frac{\mathbf{\bigtriangledown}\sigma(\mathbf{r}%
,t)}{m}.\label{8}%
\end{equation}
The momentum field can also be derived \cite{hileycallaghan} without
employing the polar form (\ref{5}) as a component of the
energy-momentum tensor of the Schr\"{o}dinger field\footnote{This is
particularly appealing when generalizations are considered to
relativistic quantum fields, in particular for the interpretation of
experiments performed in the single photon regime as in the results
reported in Ref. \cite{kocsis}, in which the observed Bohmian-like
average trajectories can be interpreted as the energy flux of the
energy-momentum tensor (just like the Poynting vector appears in the
stress-energy tensor for classical Maxwell fields).}.

Applying $\triangledown$ to Eq. (\ref{6}) and using Eq. (\ref{8}) leads to%
\begin{equation}
m\frac{d\mathbf{v}}{dt}=-\triangledown(V+Q),\label{e7}%
\end{equation}
a Newtonian-like\ law of motion.\ This justifies, in the de Broglie-Bohm
formulation, the assumption that the streamlines of the probability flow are
actually trajectories taken by a point-like particle governed by Eq.
(\ref{e7}), where the dynamics is determined not by the sole usual potential
$V$ but by the a total potential function $V+Q$ thus including a
wavefunction-dependent "quantum potential" term.

\begin{figure}[tb]
\includegraphics[height=12cm]{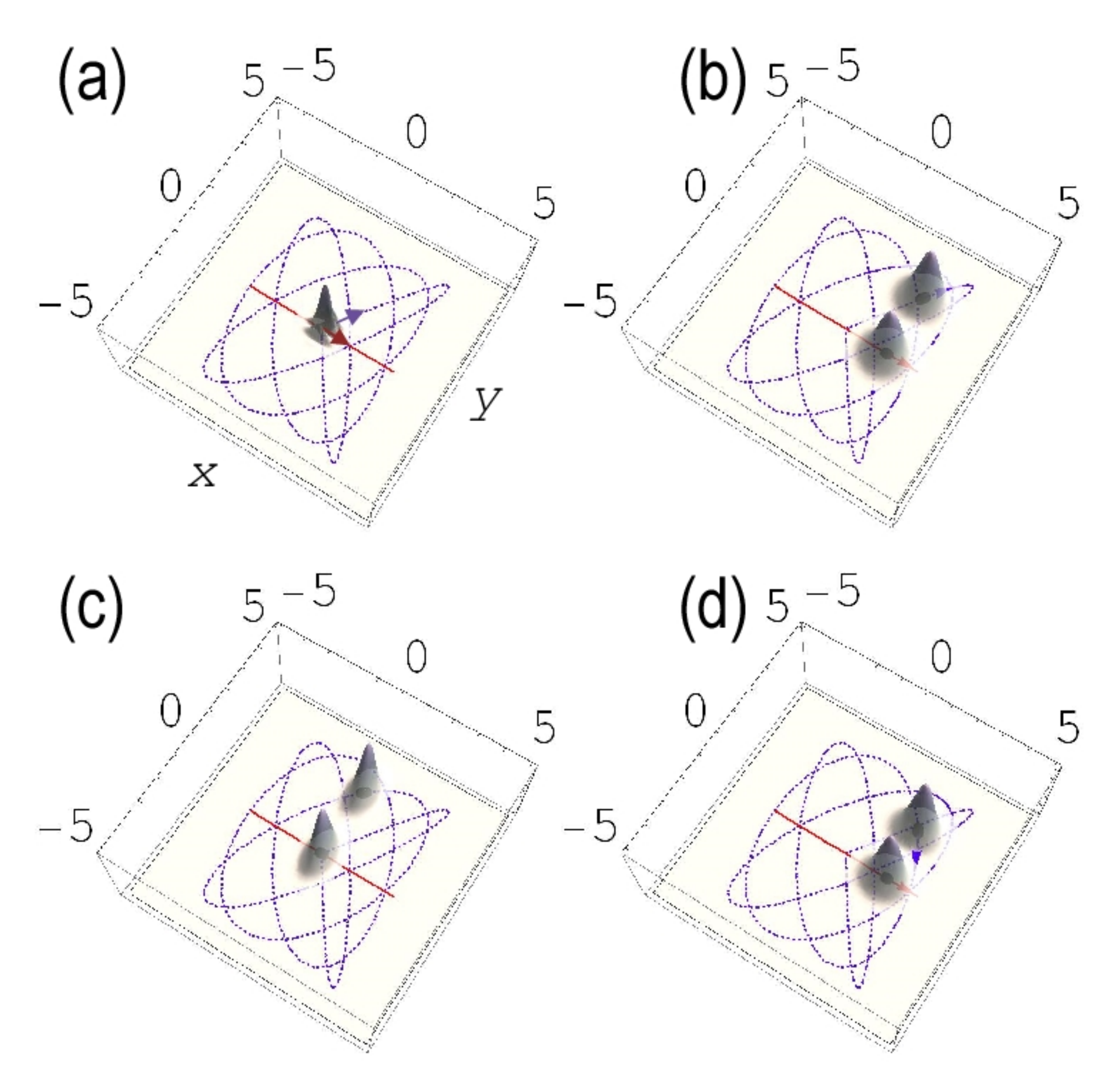}\caption{ (a) The probability density
of the initial $(t_{0}=0)$ wavefunction Eq. (\ref{10}) with $\mathbf{q}_{0}%
=0$, $J=1,2$ and $a_{1}=a_{2}$ is seen to be localized at the origin; the
direction of $\mathbf{p}_{0}^{J}$ is schematically indicated by the
arrows.\ The two classical trajectories with initial conditions $(\mathbf{q}%
_{0},\mathbf{p}_{0}^{1})$ and $(\mathbf{q}_{0},\mathbf{p}_{0}^{2})$ are shown
in dashed (purple) and solid (red) lines. Both trajectories are closed at the
origin. The probability density is then shown (b): at $t>t_{0}$ when the
wavepackets start moving; each wavepacket follows the classical guiding
trajectory $J=1,2 $; (c) at $t=t_{rec(2)},$ corresponding to the second peak
of the autocorrelation (see text and Fig.\ 3); (d) just before the two
wavepackets cross and interfere. Atomic units are used throughout, with a TDLO
having unit mass.}%
\label{fig1}%
\end{figure}

When the wavefunction vanishes the quantum potential becomes singular and
dominates the dynamics. Therefore generically Bohmian trajectories cannot be
classical \cite{mismatch}, even in semiclassical
systems.\ Nevertheless the signatures of the underlying classical dynamics in
quantum systems, that appear as large scale structures in the quantum
properties, are recovered on a statistical basis.\ An illustration is given
immediately below. The apparent conflicting situation between the classical
trajectories resulting from the path integral approach and the de Broglie-Bohm
trajectories with regards to the interpretation of the dynamics of a system
such as the TDLO becomes particularly acute in the context of weak
measurements, as illustrated in Sec. \ref{secwmt}.

\subsection{Illustration: Recurrence spectrum and returning trajectories in
the TDLO\label{illus}}

As an illustration of the different dynamical pictures that arise depending on
the nature of trajectories that are implemented, let us take our 2D
time-dependent linear oscillator (TDLO). Consider an initial wavefunction
given by Eq. (\ref{10}) with $\mathbf{q}_{0}=0$, $J=1,2$ and $a_{1}=a_{2}$
\footnote{In the examples given in this work atomic units are used throughout,
with a TDLO having unit mass.} ; the direction of $\mathbf{p}_{0}^{J}$ is
shown schematically in Fig.\ \ref{fig1}(a) along with the resulting classical
trajectories. Snapshots of the time-dependent wavepackets are shown in Fig.
\ref{fig1}(a)-(d). Each wavepacket follows the guiding trajectory
$\mathbf{q}^{J}(t).$ Both of these guiding trajectories are periodic with
period $4\pi$ (the plots in Fig.\ \ref{fig1} show the trajectories in the
interval $[0,2\pi]$).

\begin{figure}[tb]
\includegraphics[height=7cm]{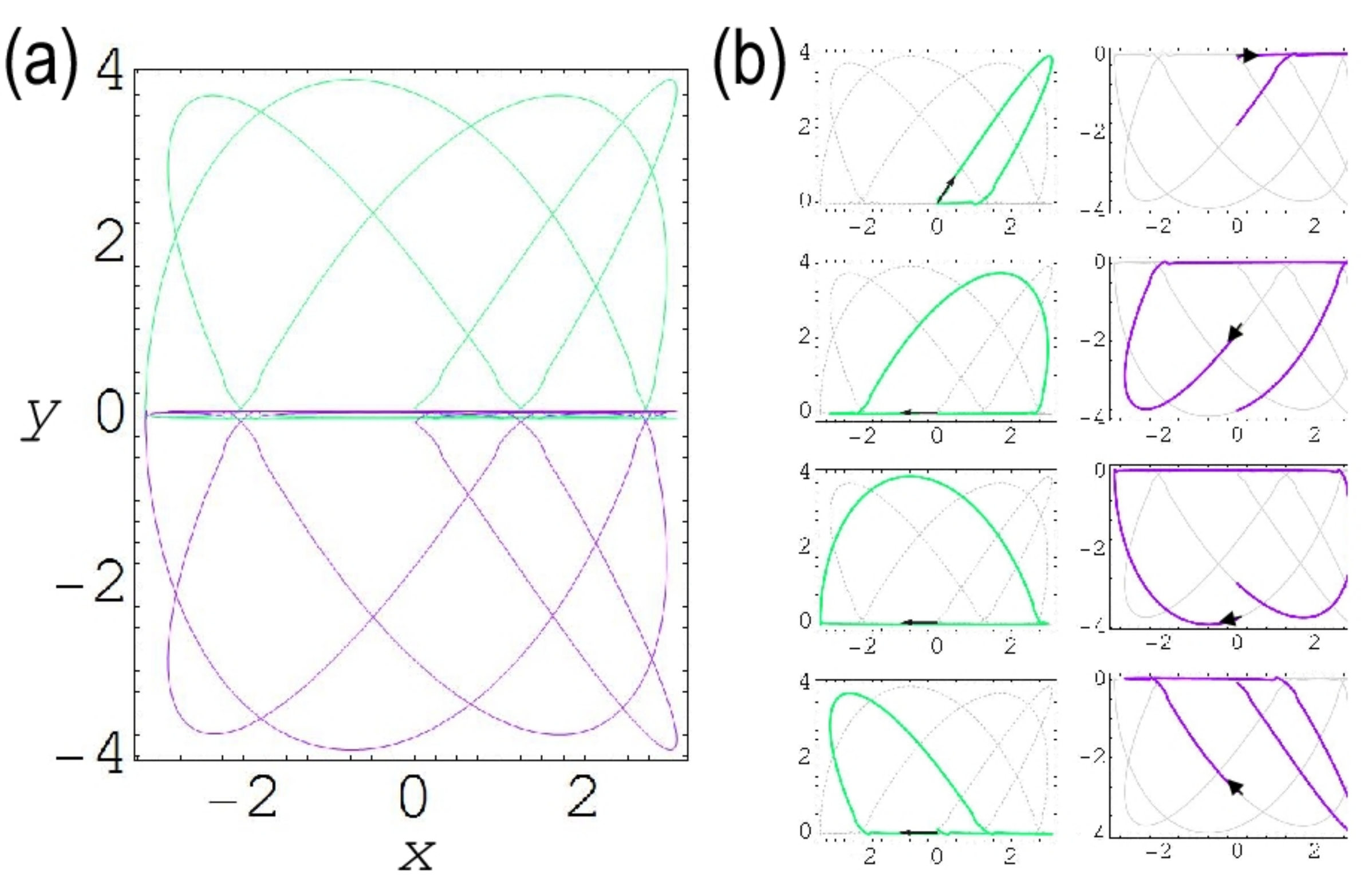}\caption{ (a) Two Bohmian trajectories
corresponding to the wavefunction shown in Fig.\ 1 are plotted in light grey
(green) and dark grey (purple); their initial conditions are near the maximum
of the wavefunction, $\mathbf{x}_{dBB}(0)=(0.01,0.09)$ and $(0.01,-0.08)$
respectively. (b): The time evolution of these 2\ Bohmian trajectories are
depicted in the time intervals (from top to bottom): $(0,t_{rec(1)})$,
$(t_{rec(1)},t_{rec(3)}),$ $(t_{rec(3)},t_{rec(5)}),$ $(t_{rec(5)}%
,t_{rec(7)}),$ where $t_{rec(j)}$ is the recurrence time defined by the return
of a wavepacket to the origin (see text for details) and $t_{rec(7)}=2\pi$.
The arrows indicate the direction of the Bohmian particle motion at the
beginning of each time interval.}%
\label{fig2}%
\end{figure}

Two typical Bohmian trajectories are shown in Fig. \ref{fig2}. The first
(resp. second) Bohmian trajectory was chosen so that shortly after $t=0,$ the
Bohmian particle sits at the maximum of the $J=1$(resp.\ $J=2$) wavepacket.
The main characteristic of the Bohmian trajectories is that they seem to
"jump" from one guiding trajectory to the other each time the wavepackets
cross or interfere. This is a simple consequence of Eq. (\ref{8}): the
velocity of the Bohmian "particle" is proportional to the overall current
density resulting from the interfering wavepackets, and by definition the
current density lines do not cross each other. Although these two families of
Bohmian trajectories are different from the classical guiding trajectories, on
a statistical basis the motion of the wavepackets along the guiding
trajectories is recovered.

\begin{figure}[t]
\includegraphics[height=6cm]{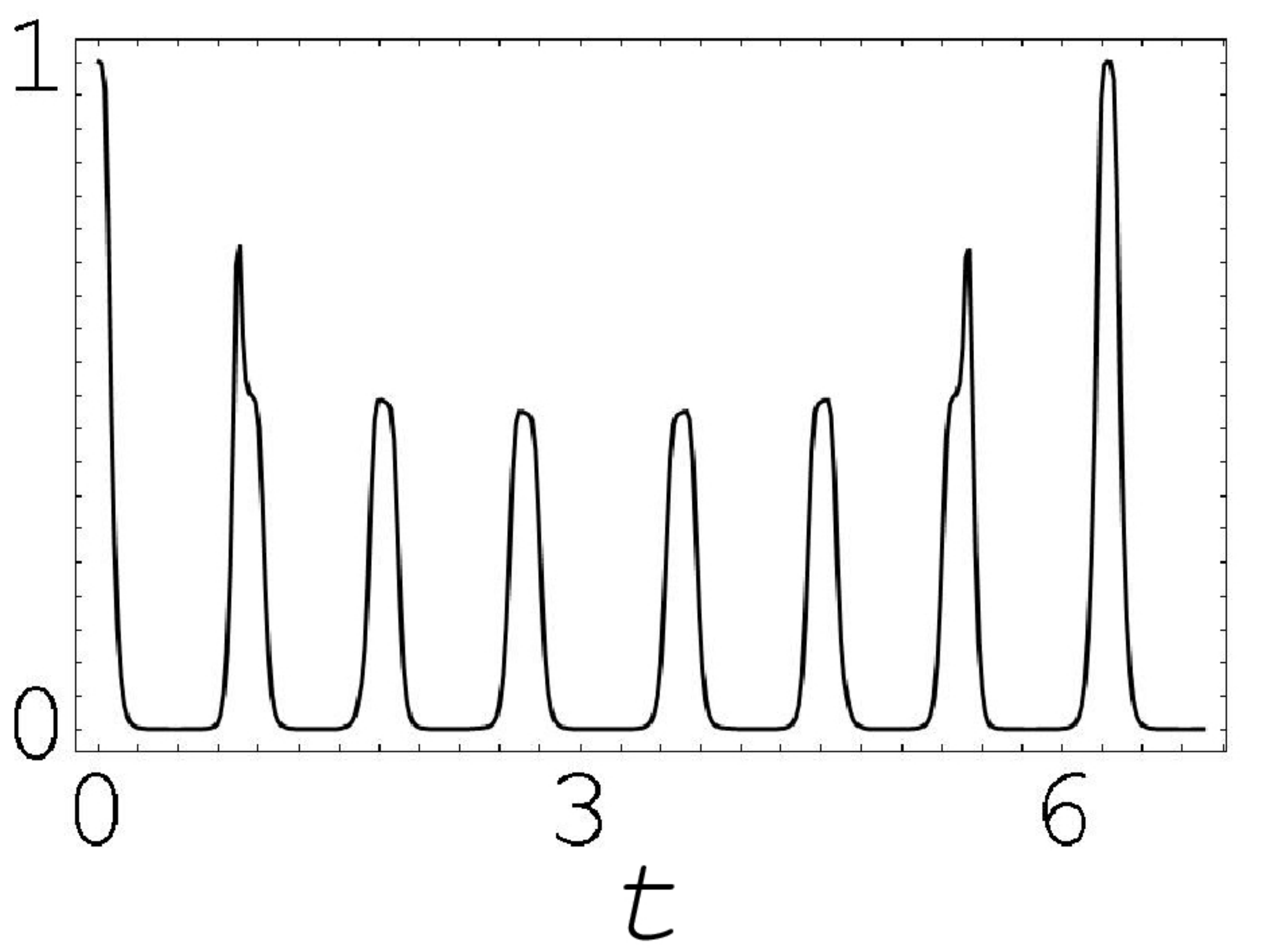}
\caption{The recurrence spectrum defined
in the text is given as a function of time. A peak occurs when part of the
wavefunction returns to the region around the origin. 7 recurrence peaks are
visible in the plot, the recurrences taking place at times $t_{rec(j)},
j=1-7$.}%
\label{fig3}%
\end{figure}

The Bohmian and classical trajectories can be seen as two alternative manners
of defining the propagation and transport of the probability density in this
TDLO. For example if one focuses on the probability density in a small region
$\mathcal{V}$ around the origin, the so-called recurrence spectrum
(Fig.\ \ref{fig3}), defined by the probability $P(t)=\int_{\mathcal{V}%
}\left\vert \psi(\mathbf{x},t)d\mathbf{x}\right\vert $ displays sharp peaks at
specific times $t_{rec}$.\ These values of $t_{rec}$ correspond to the passage
of a wavepacket in the region $\mathcal{V}$ and they are obviously given by
the times at which one of the classical guiding trajectories passes through
the origin. It can be seen from Figs. \ref{fig3} and \ref{fig1}, that the
first recurrence taking place at time $t_{rec(1)}$ is due to the wavepacket
propagating along the $J=2$ (red solid) guiding trajectory.\ Actually the
first return to the origin of the $J=1$ (dashed purple) trajectory only
happens at $t=t_{rec(7)}=2\pi$. In terms of the Bohmian trajectories, the
interpretation is more involved: as can be seen from Fig. \ \ref{fig2}(b) the
Bohmian trajectory plotted in green (light gray) passes through the origin at
$t=t_{rec(1)},t_{rec(3)},t_{rec(5)}$ and $t_{rec(7)}=2\pi$ and therefore
contributes to the relevant peaks in the recurrence spectrum of Fig.\ 3. On
the other hand the purple (dark gray) trajectory is not near the origin at
those recurrence times, but instead contributes to the peaks at the recurrence
times $t=t_{rec(2)},t_{rec(4)},t_{rec(6)}$ and $t_{rec(7)}$ (this is not shown
in the figures). We thus see that the dynamical interpretation of the
recurrence spectrum in terms of trajectories is quite different if couched in
terms of classical trajectories or given in the de Broglie-Bohm framework.

\section{Weak Measurements and trajectories in the TDLO\label{secwmt}}

\subsection{General Remarks}

We give in this section derivations and specific computations for weak
measurement of trajectories for the TDLO. We have seen in Sec.\ II\ that weak
measurements of the position and momentum observables lead to different type
of trajectories.\ A crucial difference in the protocols is that the "weak
trajectories" are obtained from the analysis of \emph{several} weak
measurement apparata interacting with the system as it evolves from a given
pre-selected state to a unique final post-selected state.\ Post-selection is
not made after each weak measurement, but only at the end of the
evolution.\ Instead the local average velocity inferred from the weak
measurement of the momentum operator is obtained by performing a \emph{single}
weak measurement immediately followed by post-selection; but in order to
obtain the velocity field, such weak measurements must be repeated by scanning
the post-selected state over the spatial region of interest.

Note also that although the pre-selected state is the initial state of the
system in both cases, the post-selected states will typically be different.
For "weak trajectories", it is useful to choose a post-selected state carrying
the dynamical information (the mean position and momentum at the time of
post-selection) of the wavepacket.\ The weak measurement of Bohmian
trajectories relies instead on post-selecting ideally to an eigenstate of the
position operator (see however Ref. \cite{robust}, in which a protocol
allowing to obtain an approximate weak measurement of the velocity field in
non-ideal conditions is presented).

\subsection{Weak trajectories and sum over paths}

\subsubsection{Weak trajectories and the underlying classical dynamics}

Let us start by specializing the weak position weasurements (\ref{wvwt}) to
the case of the TDLO with some additional assumptions. First let us take the
post-selected state to be the Gaussian%
\begin{equation}
\chi_{\mathbf{r}_{f},\mathbf{p}_{f}}(\mathbf{r},t_{f})=\left(  \frac{2}%
{\pi\delta_{f}^{2}}\right)  ^{1/2}e^{-\left(  \mathbf{r}-\mathbf{r}%
_{f}\right)  ^{2}/\delta_{f}^{2}}e^{i\mathbf{p}_{f}\cdot\left(  \mathbf{r}%
-\mathbf{r}_{f}\right)  /\hslash}.\label{25}%
\end{equation}
Recall from Eq. (\ref{wvwt}) that the expression of the weak position at time
$t_{k},$ $\left\langle \mathbf{r}(t_{k})\right\rangle _{W}$ involves the
post-selected state $\left\vert \chi(t_{k})\right\rangle $ at time $t_{k} $
which is the wavefunction $\chi_{\mathbf{r}_{f},\mathbf{p}_{f}}(\mathbf{r}%
,t_{f})$ evolved backward in time. For the TDLO, this means finding the
guiding trajectory $\mathbf{q}_{f}(t)$ having the final boundary condition
$\mathbf{q}_{f}(t_{f})=\mathbf{r}_{f}$ and $\mathbf{p}_{f}(t_{f}%
)=\mathbf{p}_{f}.$ At time $t_{k}$ the backward evolved guiding trajectory
will be found at the position $\mathbf{q}_{f}(t_{k}).$ Hence a non-vanishing
weak value will be registered by a WMA positioned near $\mathbf{R}%
^{0}\mathbf{\approx q}_{f}(t_{k})$ provided the wavefunction $\psi
^{(\mathbf{q}_{0},\mathbf{p}_{0})}(\mathbf{r},t=t_{k})$ has a substantial
amplitude in the region $\mathbf{r\approx q}_{f}(t_{k}).$

\begin{figure}[tb]
\includegraphics[height=9cm]{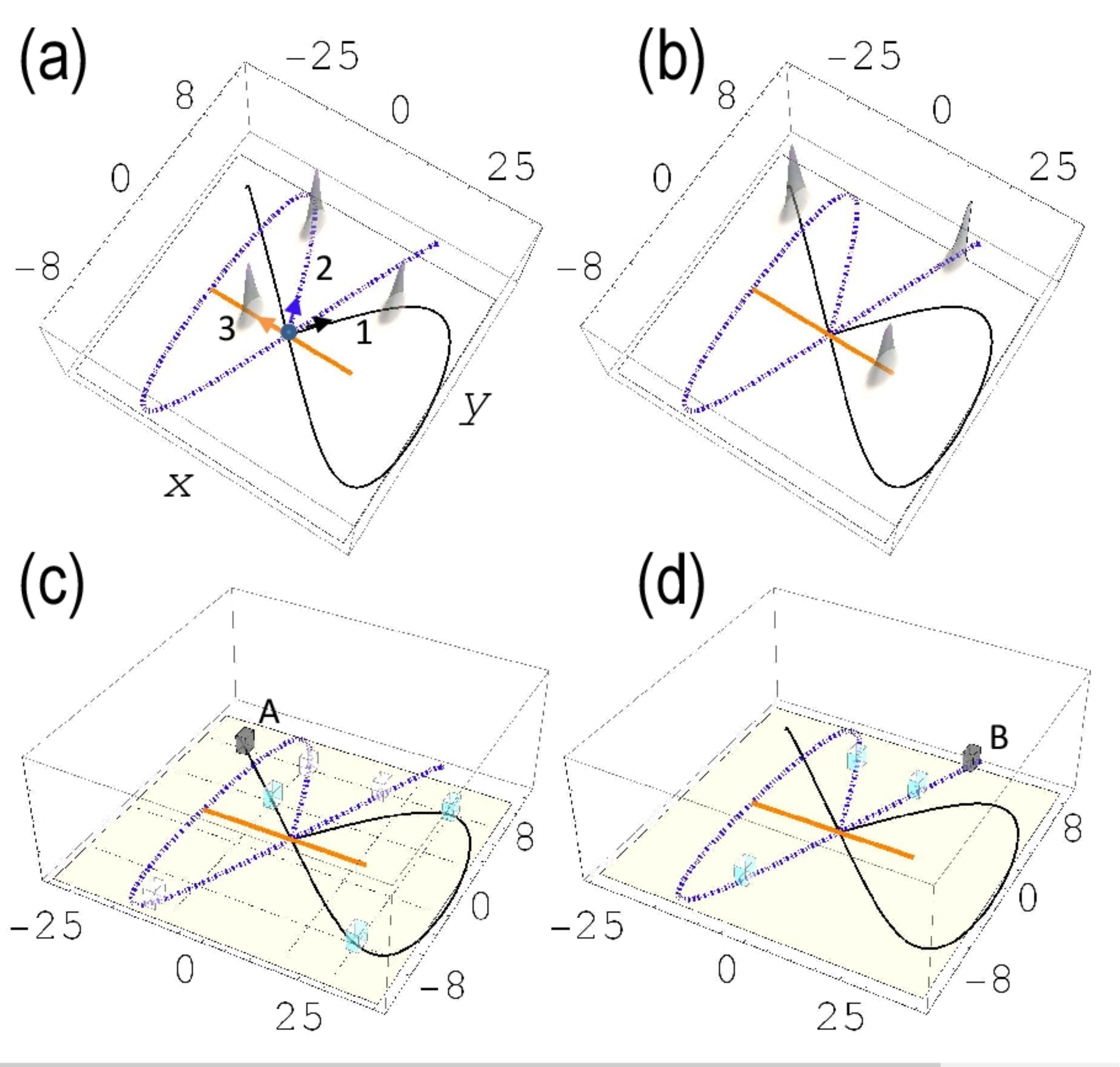}\caption{ (a) A wavefunction of the form
given by Eq. (\ref{10}) with $\mathbf{q}_{0}=0$, $J=1,2,3$, $a_{1}\approx
a_{2}\approx a_{3},$ and with $\mathbf{p}_{0}^{J}$ schematically indicated by
the arrows, initially $(t_{0}=0)$ compactly localized at the origin (blue dot)
propagates along the the 3\ guiding trajectories $J$ (see labels in the
plot).\ The probability density is shown at $t>t_{0}$ when the wavepackets
start moving along the guiding trajectories. (b) The probability density is
shown at $t=t_{f},$ the time at which the postselection is made. (c)
Postselection, represented by the gray box, is made at point $A$ along the
$J=1$ trajectory with a postselection to state (\ref{wt condi}) with $J=1$. A
grid of WMAs (weak meaurement apparata, acting as quantum probes) as defined
in the text is suggested, some WMAs being explicitly pictured (white boxes).
Only the WMAs along trajectory $A$ have their quantum state modified
(indicated by a blue shading) whereas the quantum states of the other WMAs on
the grid are left unchanged. The weak trajectory can be inferred by reading
the states of the WMAs. (d) Same as (c) when postselection is made at point
$B$ along the $J=2$ trajectory with a postselected state (\ref{wt condi}) with
$J=2$: only the WMAs along trajectory $B$ (shown by a blue shading) have their
quantum state modified, with a phase term proportional to the weak value
(\ref{wvgt}).}%
\label{fig wt-guiding}%
\end{figure}

A particular case of practical interest arises when
\begin{equation}
\chi_{\mathbf{r}_{f},\mathbf{p}_{f}}(\mathbf{r},t_{f})\mathbf{\approx}%
\psi^{(\mathbf{q}_{0},\mathbf{p}_{0}^{J})}(\mathbf{r},t_{f}),\label{pscondi}%
\end{equation}
where following the notation of Eqs. (\ref{10})-(\ref{12}) $\psi
^{(\mathbf{q}_{0},\mathbf{p}_{0}^{J})}(\mathbf{r},t_{f})$ represents the
branch of the pre-selected wavefunction $\psi(\mathbf{r},t_{0})=\sum_{J}%
a_{J}\psi^{(\mathbf{q}_{0},\mathbf{p}_{0}^{J})}(\mathbf{r},t_{0})$ [Eq.
(\ref{10})] propagating along a guiding trajectory $\mathbf{q}^{J}(t).$ Then
only a WMA positioned along the guiding trajectory $\mathbf{q}^{J}(t)$ will
display non-vanishing weak values.\ Moreover since Eq. (\ref{pscondi}) then
holds for any $t,$ each weak value $\left\langle \mathbf{r}(t_{k}%
)\right\rangle _{W}$ becomes an expectation value of the position in the state
$\psi^{(\mathbf{q}_{0},\mathbf{p}_{0}^{J})}(\mathbf{r},t_{k}),$ which is
simply the real term \footnote{This is not valid for the WMAs located at a
point where guiding trajectories cross.}%
\begin{equation}
\left\langle \mathbf{r}(t_{k})\right\rangle _{W}=\mathbf{q}^{J}(t_{k}%
).\label{wvgt}%
\end{equation}
The corresponding weak trajectory (\ref{wt}) is hence the guiding trajectory
$\mathbf{q}^{J}(t)$
\begin{equation}
\text{WT}_{\psi(t_{0}),\chi(t_{f})}=\{t_{k},\operatorname{Re}\left\langle
\mathbf{r}(t_{k})\right\rangle _{W}\}=\left\{  t_{k},\mathbf{q}^{J}%
(t_{k})\right\}  .\label{wt condi}%
\end{equation}
Note that in order to obtain (\ref{wt condi}), Eq. (\ref{pscondi}) is a
sufficient but non-necessary condition. In particular, it can be deduced from
Eq. (\ref{wv-spef}) given below that any $\chi_{\mathbf{r}_{f},\mathbf{p}_{f}%
}(\mathbf{r},t_{f})$ such that $\mathbf{q}^{J}(t_{f})=\mathbf{q}_{f}(t_{f})$
and $\mathbf{p}^{J}(t_{f})=\mathbf{p}_{f}(t_{f})$ will also lead to the weak
trajectory (\ref{wt condi}).

An illustration is given in Fig. \ref{fig wt-guiding}. An initial
wavefunction
\begin{equation}
\psi(\mathbf{r},t_{0})=\sum_{J=1}^{3}a_{J}\psi^{(\mathbf{q}_{0}=0,\mathbf{p}%
_{0}^{J})}(\mathbf{r},t_{0})\label{101}%
\end{equation}
tightly localized at the origin, subsequently expands as a sum over paths
involving 3 guiding trajectories $J=1,2,3$ [Fig. \ref{fig wt-guiding}(a)].
Post-selection will take place at $t=t_{f}$ chosen slightly after the
wavepackets have returned for the first time to the origin [the probability
density at $t=t_{f}$ is plotted in \ref{fig wt-guiding}(b)]. Let us first
choose a post-selected state given by Eq. (\ref{pscondi}) with $\mathbf{r}%
_{f}=\mathbf{r}_{A}$ [the point $A$ is displayed in Fig \ref{fig wt-guiding}%
(c)] and $J=1$. Assume a set of WMA was disposed on a grid as indicated
schematically in Fig \ref{fig wt-guiding}(c).\ Then, only the WMAs placed
along the classical trajectory $\mathbf{q}^{J=1}(t)$ record a non-vanishing
weak value. This weak value is given by Eq. (\ref{wvgt}). Fig
\ref{fig wt-guiding}(d) shows the WMA having a non-vanishing weak-value for
the same preselected state but for a postselected state obeying (\ref{pscondi}%
) at $\mathbf{r}_{f}=\mathbf{r}_{B}$ with $J=2$: those WMAs are precisely the
ones placed along the classical trajectory $\mathbf{q}^{J=2}(t)$.

\subsubsection{Isolated weak values}

$\left\langle \mathbf{r}(t_{k})\right\rangle _{W}$ can also be obtained
analytically for a post-selected state of the form (\ref{25}) but not obeying
the condition (\ref{pscondi}). Provided a WMA does not lie at positions where
different branches $\psi^{(\mathbf{q}_{0}=0,\mathbf{p}_{0}^{J})}%
(\mathbf{r},t)$ of the system wavefunction overlap, and further assuming that
the wavepackets are narrower than the range of the system-WMA interaction
given by $f\left(  \left\vert \mathbf{\hat{r}}-\mathbf{R}_{k}^{0}\right\vert
^{2}\right)  $, the integrals in Eq (\ref{wvwt}) defining the weak value can
be computed analytically and where non-vanishing put in the compact form%
\begin{equation}
\left\langle \mathbf{r}(t_{k})\right\rangle _{W}=\mathbf{q}^{J_{k}}%
(t_{k})\mathbf{+}\left[  \mathbf{q}^{J_{k}}(t_{k})-\mathbf{q}_{f}%
(t_{k})\right]  \cdot\mathbf{m}_{1}(t_{k})+\left[  \mathbf{p}^{J_{k}}%
(t_{k})-\mathbf{p}_{f}(t_{k})\right]  \cdot\mathbf{m}_{2}(t_{k}%
).\label{wv-spef}%
\end{equation}
As above $\mathbf{q}^{J_{k}}(t_{k})$ and $\mathbf{p}^{J_{k}}(t_{k})$ are the
position and momentum of the classical guiding trajectory $J_{k}$ at the time
the system interacts with the $k$th WMA. $\mathbf{q}_{f}(t_{k})$ and
$\mathbf{p}_{f}(t_{k})$ are the position and momentum at time $t_{k}$ of the
trajectory having the boundary conditions fixed by the choice of
post-selection, $\mathbf{q}_{f}(t_{f})=\mathbf{r}_{f}$ and $\mathbf{p}%
_{f}(t_{f})=\mathbf{p}_{f}$ (so that $\mathbf{q}_{f}(t)$ is formally the
trajectory with the final boundary conditions determined by postselection
evolved backward in time). $\mathbf{m}_{1}(t_{k})$ and $\mathbf{m}_{2}(t_{k})
$ are purely time-dependent \emph{complex} functions.

\begin{figure}[tb]
\includegraphics[height=6cm]{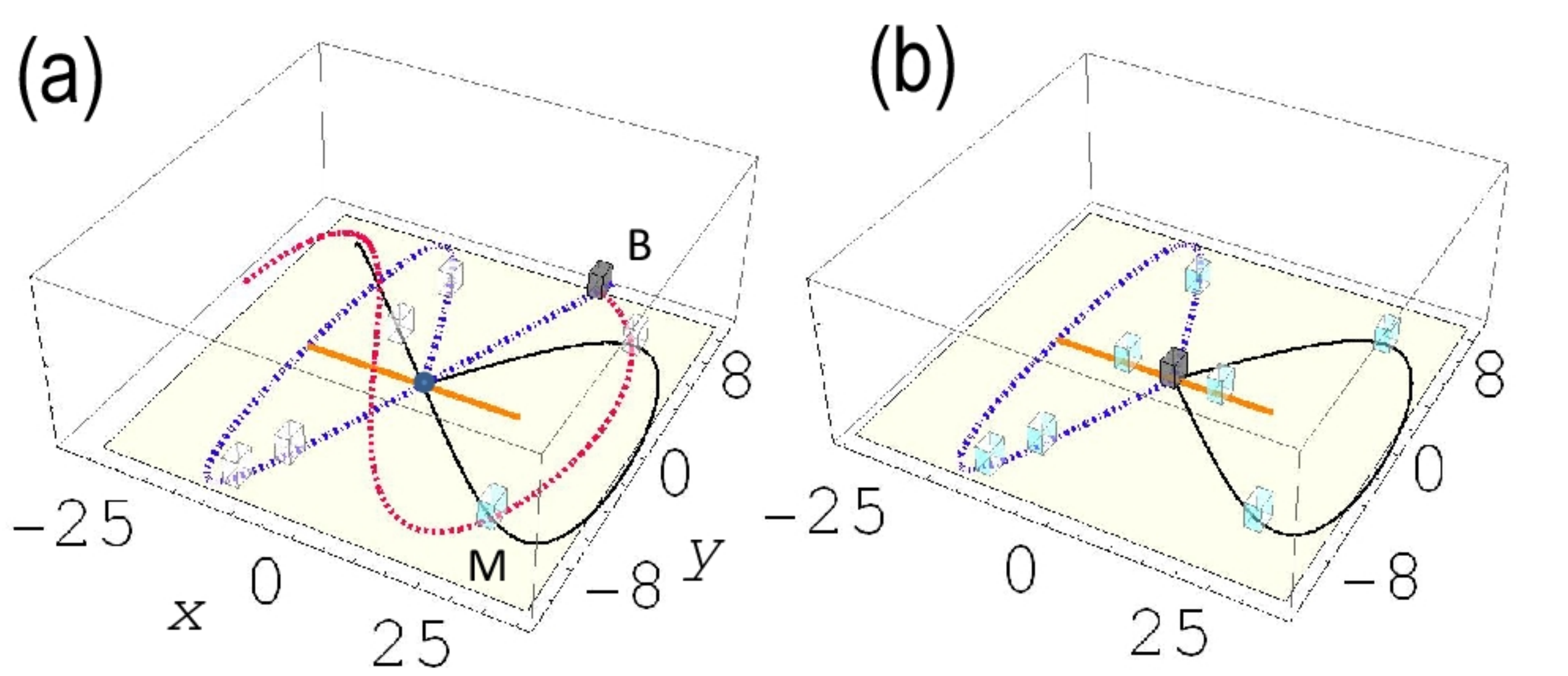}\caption{(a) \emph{Isolated weak value}
: Postselection takes place at point $B$ along the $J=2$ trajectory, but
unlike the case pictured in Fig. \ref{fig wt-guiding}(d), the postselected
state is of the form (\ref{25}) but does not obey the condition (\ref{pscondi}%
); instead the postselected momentum is ``arbitrary'', and corresponds to the
trajectory evolving backward in time shown in red (dashed grey). Then the WMAs
placed in a grid are generally not modified (a few are shown in the figure,
left unshaded) since the weak values vanish and no weak trajectory can be
defined.\ If the post-selected backward-evolved guiding trajectory crosses
accidentally a wavepacket traveling on one of the system's guiding
trajectories, the weak value will be a complex number and the relevant WMA
will be modified, as indicated by the light blue shading of the WMA labeled
$M$. This gives rise to an isolated WMA being modified, and hence no weak
trajectory can be defined. (b) \emph{Sum over paths}: Post-selection takes
place at $\mathbf{r}_{f}=0$ when the wavefunction first returns to the origin
(grey box), with a post-selected state of the form (\ref{condi feynman path}).
The WMAs along the three guiding trajectories $J=1,2,3 $ are modified (shaded
boxes) each indicating a weak value $\mathbf{q}^{J}(t_{k})$ given by
(\ref{fpwv}) (the state of the other WMAs on the grid -- not shown on the
figure -- is left unchanged). This allows to infer weak trajectories
corresponding to the sum over paths appearing in the Feynman propagator.}%
\label{fig-isolated}%
\end{figure}

Note that there is now an index $k$ at $J_{k}$ given that Eq. (\ref{pscondi})
is not obeyed.\ In that case if postselection at $t_{f}$ happens on, a branch,
say $J_{f}$ of the wavefunction, a nonzero weak value $\left\langle
\mathbf{r}(t_{k})\right\rangle _{W}$ can only be obtained provided the
backward evolving $\mathbf{q}_{f}(t)$ crosses "accidentally", at some time
$t_{k}$ a wavepacket moving along another branch, denoted $J_{k}$ (thereby
evolving along the classical trajectory $\mathbf{q}^{J_{k}}(t)$). At some
other time $t_{k^{\prime}}$, one can imagine that a different trajectory
$\mathbf{q}^{J_{k^{\prime}}}(t)$ may be crossed, especially if the inital
wavefunction of the form (\ref{101}) contains many branches. In this
situation, the weak values will be nonzero for a small number of isolated
points. Compared to the previous case, where a weak trajectory can be inferred
from (\ref{wt condi}) involving in principle a dense number of closely
positioned WMAs, the set $\{t_{k},\mathbf{R}_{k}^{0}\}$ containing a few
points cannot be said to form a trajectory, but rather isolated points
belonging to different branches of the wavefunction. An exemple is given in
Fig. \ref{fig-isolated}(a).

When the conditions stated above Eq. (\ref{wv-spef}) are not fulfilled, then
$\left\langle \mathbf{r}(t_{k})\right\rangle _{W}$ must be computed
numerically, though the compatibility condition for obtaining non-zero weak
values (overlap of $\chi_{\mathbf{r}_{f},\mathbf{p}_{f}}(\mathbf{r},t_{k})$
and $\psi(\mathbf{r},t_{0})$ in the neighborhood of $\mathbf{R}_{k}^{0}$) can
generally be inferred from the classical dynamics that determine the guiding trajectories.

\subsubsection{Sum over paths}

We have just seen that when post-selection takes place along a given branch of
the wavefunction, then if the postselected state is dynamically compatible
with the guiding classical trajectory carrying that branch, then only the WMAs
placed along that guiding trajectory will display non-zero weak values.
Post-selecting appropriately on a different branch will instead yield non-zero
weak values along the guiding trajectory associated with that specific branch.
Now since the wavefunction (\ref{101}), or more generically the semiclassical
propagator (\ref{e10}) involves a sum over paths, it would be valuable if the
corresponding weak trajectories could be detected simultaneously by the weak
measurement apparata.

This is possible if post-selection is made at some final position
$\mathbf{r}_{f}$ where two or more trajectories cross and provided the
post-selected state can be tailored to take the approximate form%
\begin{equation}
\chi_{\mathbf{r}_{f}}(\mathbf{r},t_{f})\mathbf{\approx}\sum_{K}c_{K}%
\chi_{\mathbf{r}_{f},\mathbf{p}_{f}^{K}}(\mathbf{r},t_{f}%
),\label{condi feynman path}%
\end{equation}
where the $c_{K}$ are arbitrary coefficients and the $\chi_{\mathbf{r}%
_{f},\mathbf{p}_{f}^{K}}(\mathbf{r},t_{f})\mathbf{\approx}\psi^{(\mathbf{q}%
_{0},\mathbf{p}_{0}^{K})}(\mathbf{r},t_{f})$ as in Eq. (\ref{pscondi}).
Plugging in Eqs. (\ref{12}) and (\ref{condi feynman path}) in the weak value
definitions (\ref{wv-def}) and (\ref{wvwt}) yields
\begin{align}
\left\langle \mathbf{r}(t_{k};\mathbf{R}_{k}^{0})\right\rangle _{W}  &
=\frac{\left\langle \chi_{\mathbf{r}_{f}}(t_{f})\right\vert U(t_{f}%
,t_{k})\mathbf{\hat{r}}f\left(  \left\vert \mathbf{\hat{r}}-\mathbf{R}_{k}%
^{0}\right\vert ^{2}\right)  \left[  U(t_{k},t_{0})\left\vert \psi
(t_{0})\right\rangle \right]  }{\left\langle \chi_{\mathbf{r}_{f}}%
(t_{f})\right\vert U(t_{f},t_{0})\left\vert \psi(t_{0})\right\rangle }\\
&  =\frac{\sum_{K,J}c_{K}^{\ast}a_{J}\left\langle \chi_{\mathbf{p}_{f}^{K}%
}(t_{k})\right\vert \mathbf{\hat{r}}f\left(  \left\vert \mathbf{\hat{r}%
}-\mathbf{R}_{k}^{0}\right\vert ^{2}\right)  \left\vert \psi^{(\mathbf{q}%
_{0},\mathbf{p}_{0}^{J})}(t_{k})\right\rangle }{\sum_{K,J}c_{K}^{\ast}%
a_{J}\left\langle \chi_{\mathbf{p}_{f}^{K}}(t_{k})\right\vert \left.
\psi^{(\mathbf{q}_{0},\mathbf{p}_{0}^{J})}(t_{k})\right\rangle }.\label{201}%
\end{align}
Assuming as we have done up to now that the WMAs set at places where the
different branches interfere are disregarded, a typical WMA positioned at
$\mathbf{R}_{k}^{0}$ will therefore interact at most with the wavepacket
propagating along a given branch, say $J$.\ In turn only the same branch $J$
of the postselected state (\ref{condi feynman path}) will overlap with the
system wavefunction at $\mathbf{R}_{k}^{0}$ and Eq. (\ref{201}) becomes%
\begin{equation}
\left\langle \mathbf{r}(t_{k});\mathbf{R}_{k}^{0};J\right\rangle _{W}%
=\frac{\left\langle \chi_{\mathbf{p}_{f}^{J}}(t_{k})\right\vert \mathbf{\hat
{r}}f\left(  \left\vert \mathbf{\hat{r}}-\mathbf{R}_{k}^{0}\right\vert
^{2}\right)  \left\vert \psi^{(\mathbf{q}_{0},\mathbf{p}_{0}^{J})}%
(t_{k})\right\rangle }{\left\langle \chi_{\mathbf{p}_{f}^{J}}(t_{k}%
)\right\vert \left.  \psi^{(\mathbf{q}_{0},\mathbf{p}_{0}^{J})}(t_{k}%
)\right\rangle }=\mathbf{q}^{J}(t_{k});\label{fpwv}%
\end{equation}
it is now necessary to explictly state the branch $J$ relevant to the weak
value.\ Indeed, possibly at the same time a different WMA positioned at
$\mathbf{R}_{k^{\prime}}^{0}$ will have interacted with another branch
$J^{\prime}$ of the system wavefunction consistent with the postselection
condition (\ref{condi feynman path}), recording the weak value $\left\langle
\mathbf{r}(t_{k});\mathbf{R}_{k^{\prime}}^{0};J^{\prime}\right\rangle _{W}.$
If there is a sufficient number of WMAs it is then straightforward to arrange
the non-zero weak values extracted from the weak measurement apparata into
time-ordered sets corresponding to different trajectories%
\begin{align}
\text{WT}_{\psi(t_{0}),\chi(t_{f})}(J_{1})  &  =\{t_{k},\operatorname{Re}%
\left\langle \mathbf{r}(t_{k});\mathbf{R}_{k}^{0};J_{1}\right\rangle
_{W}\}=\left\{  t_{k},\mathbf{q}^{J_{1}}(t_{k})\right\} \\
\text{WT}_{\psi(t_{0}),\chi(t_{f})}(J_{2})  &  =\{t_{k},\operatorname{Re}%
\left\langle \mathbf{r}(t_{k});\mathbf{R}_{k}^{0};J_{2}\right\rangle
_{W}\}=\left\{  t_{k},\mathbf{q}^{J_{2}}(t_{k})\right\}  ...
\end{align}
These trajectories resulting from the weak measurements of the position are a
subset of the sum over paths constitutive of the propagating wavefunction
compatible with the postselected state (\ref{condi feynman path}).

Consider for example the situation previously shown in Fig.
\ref{fig wt-guiding} and assume postselection is made at $t_{f}=t_{r}$, when
the wavepackets return for the first time to the origin $\mathbf{r}_{f}%
=0$.\ The situation is represented in Fig. \ref{fig-isolated}(b):
assuming a set of weakly interacting measurement apparata (WMA) laid
out in a grid, the WMAs placed along the 3 guiding trajectories
$J=1,$ 2 and $3$ will have their quantum state modified according to
the weak value $\mathbf{q}^{J}(t_{k})$. By retrieving the weak
values, a set of weak trajectories corresponding to the three
classical guiding trajectories along which the wavepackets move can
be defined. This shows that by postselecting appropriately at a
position where several Feynman paths cross, it is in principle
possible to observe the sum over paths as weak trajectories
resulting from the interaction between WMAs and the system
wavefunction.

\subsection{Weak measurement of the velocity field and Bohmian trajectories}

\begin{figure}[tb]
\includegraphics[height=7cm]{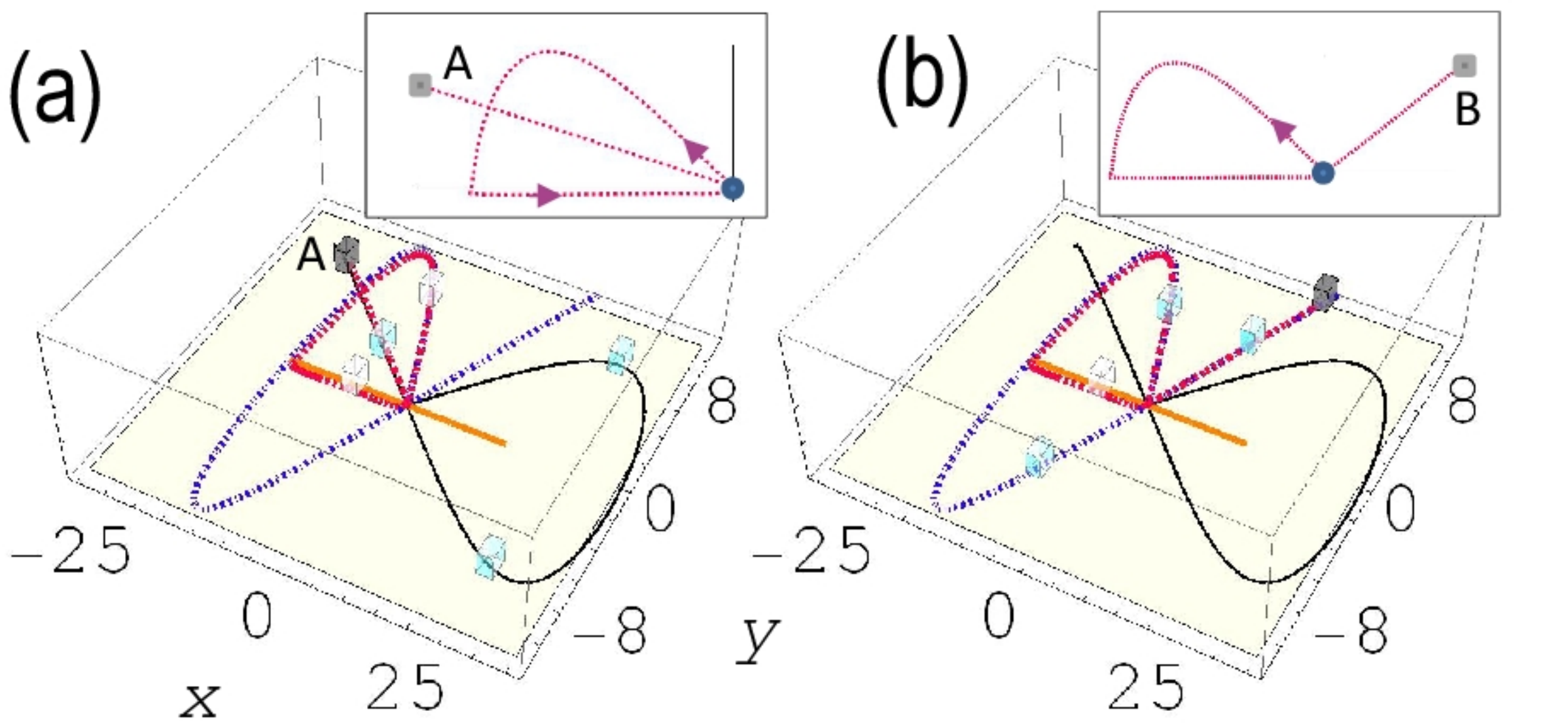}\caption{(a) A typical Bohmian
trajectory with its final ($t=t_{f})$position at the postselection point $A$
(on the $J=1\ $guiding trajectory) is shown (dark grey [online:red] thick
dashed line). The postselection condition is the one illustrated in Fig
\ref{fig wt-guiding}(c), for which the WMAs that have interacted (shaded in
the figure) are only along the $J=1$ trajectory (all the other WMAs, including
those along the Bohmian trajectory, are left unshaded). The inset details the
Bohmian trajectory, starting at the origin, first following the wavepacket
travelling along the $J=2$ guiding trajectory then jumping to the wavepacket
along the $J=3$ guiding trajectory, going back to the origin and leaving along
the wavepacket following the $J=1\ $trajectory. All the Bohmian trajectories
with a final position in the vicinity of $A$ have the form shown in the inset.
(b): Same as (a) but for a postselection condition at $t=t_{f}$ at point $B,$
corresponding to the example shown in Fig \ref{fig wt-guiding}(d). A typical
Bohmian trajectory having its endpoint in the vicinity of $B$ is shown [dark
grey (online:red) thick dashed line]. Here the shaded WMAs are those along the
$J=2$ guiding trajecotry. The inset details how the Bohmian trajectory unfolds
in time.}%
\label{bohmian}%
\end{figure}

As we have seen above [Eqs. (\ref{wm-mom})-(\ref{wm-vel})], the weak
measurement of the momentum operator followed immediately by postselection to
an eigenstate $\left\vert \mathbf{r}_{f}\right\rangle $ of the position
operator yields a velocity field that turns out to correspond to the local
velocity at $\mathbf{r}_{f}$ of the particle constitutive of the Bohmian model
characterized by the law of motion (\ref{e7}). Here each weak measurement,
made by a single WMA positioned at $\mathbf{R}^{0}=\mathbf{r}_{f},$ is
followed by a projective measurement ideally at the same position.\ The
projective measurement terminates the system evolution, so the procedure must
be repeated first for the same $t$ throughout space (or at least where the
wavefunction amplitude is known to be non-negligible) and this must be done
again for each value of $t$ under consideration. Overall, these weak
measurements allow to map unambiguously the velocity field $\mathbf{v}%
(\mathbf{r}_{f},t)=\operatorname{Re}\left\langle \mathbf{p}(t)\right\rangle
_{W}/m$ at each space-time point.

Note that strictly speaking, it is not possible to deduce unambiguously
particle trajectories from a finite sample of velocity field values
$\mathbf{v}(\mathbf{r}_{f},t_{k})$, without assuming in the first place that
the streamlines correspond to the actual motion of particles. The additional
specific assumption (\ref{e7}) needs to be made. Then the Bohmian trajectories
can be integrated from the velocity field. The Bohmian trajectories shown in
Fig. \ref{bohmian} have been computed by numerical integration. They
correspond to the wavefunction displayed in Fig. \ref{fig wt-guiding}.

Fig. \ref{bohmian}(a) shows a typical Bohmian trajectory consistent with
postselection at point $A$ [see Fig \ref{fig wt-guiding}(c)].\ In order to
infer that Bohmian trajectory from experimental observations, one would need
to repeat the weak measurement procedure of Sec. \ref{wmbt} by taking
different postselection points at different times. Instead in Fig.
\ref{bohmian}(a) we directly show a Bohmian trajectory arriving at point $A$
in the the postselection region.\ If we assume we have carried out the
postselection at $A$ as specified in Fig. 4(c),then only the WMAs along
trajectory $J=1$ will be shaded, the other WMAs on the grid do not have their
quantum state modified. The trajectory shown in Fig. \ref{bohmian}(a) is the
one that at $t=t_{f}$ has the final position $\mathbf{x}_{dBB}(t_{f}%
)=\mathbf{q}^{J=1}(t_{f})$ ie at the maximum of the wavepacket that has
followed the guiding trajectory $J=1$.\ This trajectory is ``typical'' in the
sense that all the other Bohmian trajectories having the position at $t=t_{f}$
in the vicinity of the postselected point $\mathbf{q}^{J=1}(t_{f})$ have the
same topology, leaving initially the origin along with the wavepacket
travelling along $J=2,$ then jumping to the wavepacket along $J=3, $ going
back to the origin then leaving the origin along the wavepacket following the
$J=1\ $trajectory [see the inset in Fig.\ \ref{bohmian}(a)]. The important
feature is that these Bohmian trajectories are unrelated to the weak
trajectory measurements: for example the WMAs on path $J=1$ in the $x>0$ plane
have interacted with the system (since their quantum state has been modified),
although the Bohmian trajectories detected at $A$ do not reach these
WMAs.\ Conversely as seen in Fig. \ref{bohmian}(a) there are WMAs that are
left unchanged (that are interpreted as not having interacted with the system
given the postselection, represented by the unshaded boxes) on the path of the
Bohmian trajectories.

Fig. \ref{bohmian}(b) illustrates the same type of situation shown in Fig.
\ref{bohmian}(a) but for a weak trajectory postselected at point $B$ under the
conditions depicted in Fig \ref{fig wt-guiding}(d). The typical Bohmian
trajectory displayed in the figure ending at $B$ at the postselection time
$t=t_{f}$, is not related to the shaded and unshaded WMAs, having interacted
(or not) with the system along the weak trajectories.

\section{Discussion and Conclusion}

The starting point of this paper was to note that when trajectories
are employed to interpret the dynamics of quantum systems, the
classical trajectories of the semiclassical path integral propagator
on the one hand and the de Broglie Bohm trajectories on the other
give rise to different accounts of the dynamics taking place in the
system. While this is not a problem if these types of trajectories
are envisaged as computational tools or mathematical artifacts, the
main idea developed in this work was to introduce weak measurements
in order to implement a practically non-disturbing observational
windows that would allow to follow the evolution of the system.\
This was done by employing a model, a 2D time-dependent linear
oscillator, introduced as a manner of simulating a more complex (but
less tractable) system in the semiclassical regime.

We have seen that the classical Feynman trajectories can be observed by weakly
measuring the \emph{position} of the system by an array of weakly interacting
devices, followed by a single postselection. Taking an arbitrary postselection
state does not yield weak trajectories, given that then none of the WMAs will
show signs of interaction with the system (their quantum state has not been
modified, except accidentally, giving rise to an isolated weak value). The
dynamical compatibility condition requires that the backward postselected
state overlaps with a trajectory of the propagator.\ Only then will the entire
set of WMAs along that trajectory indicate they have interacted with the
system (their quantum state being modified by the corresponding weak value).
In this sense, the postselected state must contain the information on the
Feynman path appearing as a weak trajectory. Alternatively, a trial and error
procedure sampling the parameter space for the postselection state can be
employed, monitoring the states of the WMAs until they indicate a continuous
trajectory. This will precisely be the weak trajectory of a classical Feynman
propagator path.

We also saw that Bohmian trajectories can be inferred from the weak
measurement of \emph{momentum}.  This is particularly noteworthy,
given that the Bohmian momentum is not directly connected to the
eigenvalues of the momentum operator. Several weak measurements must
be done by sampling all the spatial domain, and this must be
repeated at each time in order to follow the evolution of the flow.
Defining Bohmian trajectories from the velocity requires an
additional assumption, associating the motion of particle with the
lines of the current density. In this sense, the weak measurements
do not directly yield the observation of the Bohmian trajectories,
but that of the momentum field.

As pointed out above [Eq. (\ref{av})], the weak measurement of
momentum usually relies on inferring a velocity from the difference
between two position measurements. In practice it is not possible to
do the position measurements with infinite precision, and it has
recently \cite{robust} been suggested to model the finite resolution
using the POVM (Positive Operator valued Measure) framework. The
authors of Ref. \cite{robust} have in particular shown under which
conditions position Gaussian measurement operators allow to recover
the Bohmian field velocity. In the context of weak trajectories,
post-selection to an ideally well-known position eigenstate is not
required, but Gaussian measurement operators could be useful to
model post-selection to a state having the characteristics of the
final wavepacket  $\left\vert \chi(t_{f})\right\rangle $.

So can weak measurements open an observational window that would
allow to give a ``correct'' account of the dynamics of a quantum
system in terms of trajectories? There are certainly different
answers that can be given to this question, ultimately relying on
interpretational commitments (including the meaning of the weak
values). A consensual option (having a Copenhagen interpretation
flavour) would state that the type of trajectories that one sees
depends on the context -- the type of weak measurements that are
being made and hence the entire experimental setup including the
WMAs (and in particular in their interaction with the system); such
an answer would obviously undermine the idea that there is a
meaningful underlying dynamics that can be understood in terms of
trajectories, as these would be relegated to being artifacts or
computational tools.

Still, the fact that weak trajectories are obtained from a series of
weak measurements of positions given a final postselected state, as
opposed to being inferred from a mean velocity field at each point,
is a natural feature with regard to the definition of a space-time
trajectory. Indeed, if we take the weak values in the original
\cite{AAV} sense as referring to a generalized value for an
observable obtained without appreciably disturbing the system
evolution, the weak values of the position for a chosen
postselection suffice in order to extract a trajectory. On the other
hand the observation of the momentum weak value at a postselected
space-time point does not as such allow to observe a trajectory, but
the local weak value of the momentum field.

Moreover, sticking to the situation illustrated in Fig.\
\ref{bohmian}(a), it is noticeable that there is no correlation
between the Bohmian trajectories detected at the postselection point
$A$ and the WMAs that interacted with the system (along parts of the
$J=1$ guiding trajectory in which there are no Bohmian trajectories
reaching $A$) and those that haven't (the Bohmian trajectories do
not trigger the WMAs along part of their route although the
interaction involves the position). This fact could apparently be
taken as an argument against the relevance of interpreting the
dynamics in terms of Bohmian trajectories, on the ground that the
particle does not comply with its role (which is to make detectors
click).\ This is not the case, however: one should indeed bear in
mind that the weak interactions are unitary and do not translate as
clicks until the WMAs themselves are measured.\ In addition, given
the non-local character of the de Broglie Bohm dynamics, the
internal states of the WMAs must be taken into account explicitly,
and this may affect non-locally some features of the Bohmian
dynamics, in particular the "no-crossing" rule \cite{hiley2006}.
While it is well-known that Bohmian trajectories are modified in
genuine open systems relative to the Bohmian trajectories of the
closed system (the ones we were interested in throughout this work),
the extent to which this aspect subsists in the case of weak
measurements remains to be investigated.

\appendix

\section{ Closed form solutions for the time-dependent harmonic oscillator}

The Hamiltonian for a one dimensional linear oscillator of mass $m$ with a
time-dependent frequency $V(t)$ is given by%
\begin{equation}
H(t)=\frac{p^{2}}{2m}+\frac{m}{2}V(t)q^{2}.\label{ham}%
\end{equation}
We will not distinguish in the notation the classical Hamiltonian and
phase-space variables from the corresponding quantum Hamiltonian and operators
unless required by the context. The solutions of the Schr\"{o}dinger equation
can be obtained exactly by employing different methods, like algebraic methods
\cite{alg}, path integrals \cite{lawande} or the more popular procedure based
on solving for the eigenfunctions of dynamical invariants \cite{LR}. We will
employ a version \cite{moshinskyPRA} of the latter method involving solutions
of Ermakov systems, which is known to be well suited to working with Gaussian wavefunctions.\

The Lagrangian corresponding to the Hamiltonian (\ref{ham}) is%
\begin{equation}
\mathcal{L}=\frac{m}{2}\dot{q}^{2}-\frac{m}{2}V(t)q^{2}\label{lag}%
\end{equation}
and leads to the classical equation of motion%
\begin{equation}
\partial_{t}^{2}q(t)+V(t)q(t)=0.\label{er1}%
\end{equation}
Employing an amplitude-phase decomposition of the classical solution $q(t)$ in
the form
\begin{equation}
q(t)=\alpha(t)\left(  c_{1}\cos\left(  \phi(t)-\phi(t_{0})\right)  +c_{2}%
\sin\left(  \phi(t)-\phi(t_{0})\right)  \right) \label{ct}%
\end{equation}
where $\alpha(t)$ is the amplitude and $\phi(t)$ the phase leads to an
auxiliary nonlinear equation%
\begin{equation}
\frac{\partial_{t}^{2}\alpha(t)}{\alpha(t)}+V(t)=\frac{c_{0}^{2}}{\alpha
^{4}(t)}\label{er2}%
\end{equation}
along with the condition%
\begin{equation}
\partial_{t}\phi(t)=\frac{c_{0}}{\alpha^{2}(t)}.\label{faz}%
\end{equation}
Eqs. (\ref{er1})-(\ref{er2}) form an Ermakov system \cite{ermakov} linking the
solutions of a linear and a nonlinear equation. The $c_{i}$ in the equations
above denote real constants.

The reason for introducing Ermakov systems is that the time-dependent
Schr\"{o}dinger equation with the Hamiltonian (\ref{ham}),%
\begin{equation}
i\hbar\partial_{t}\psi(x,t)=\left[  -\frac{\hbar^{2}}{2m}\partial_{x}%
^{2}+\frac{m}{2}V(t)x^{2}\right]  \psi(x,t)\label{schro}%
\end{equation}
admits the closed form solution%
\begin{align}
\psi^{(q_{0},p_{0})}(x,t)  &  =\left(  \frac{2m}{\pi\alpha^{2}(t)}\right)
^{1/4}e^{-\left[  x-q(t)\right]  ^{2}\left(  \frac{m}{\alpha(t)^{2}}%
-\frac{im\alpha^{\prime}(t)}{2\hbar\alpha(t)}\right)  }\nonumber\\
&  e^{ip(t)\left[  x-q(t)\right]  /\hbar}e^{\frac{i}{2\hbar}\left[
p(t)q(t)-p_{0}q_{0}\right]  }e^{-\frac{i}{2}\left[  \phi(t)-\phi_{0}\right]
}.\label{psisol}%
\end{align}
Hence the exact solutions to the Schr\"{o}dinger equation can be obtained from
the knowledge of the solutions to the Ermakov system, namely $q(t)$ which is
the solution of the linear equation (that is also the classical equation of
motion) and $\alpha(t)$ (and its integral $\phi(t)$) which is a solution of
the nonlinear Ermakov equation. It can be checked explicitly by plugging in
Eq. (\ref{psisol}) into Eq. (\ref{schro}) requires setting the constant
$c_{0}$ of Eq (\ref{er2}) to $c_{0}=2\hbar$. Finally, the notation $q_{0}$
etc. indicates the values of the functions at $t=t_{0},$ ie $q_{0}\equiv
q(t_{0})$ and so on. Let us choose specifically the amplitude function such
that $\alpha^{\prime}(t_{0})=0$. Then at $t=t_{0}$ the initial wavefunction
(\ref{psisol}) is given by%
\begin{equation}
\psi^{(q_{0},p_{0})}(x,t_{0})=\left(  \frac{2m}{\pi\alpha_{0}^{2}}\right)
^{1/4}e^{-m\left[  x-q_{0}\right]  ^{2}/\alpha_{0}^{2}}e^{ip_{0}\left[
x_{0}-q_{0}\right]  /\hbar},\label{1}%
\end{equation}
that is a standard Gaussian. $q_{0}$ and $p_{0}$ that appeared as parameters
in Eq. (\ref{psisol}) thus correspond to the average initial position and
momentum of the initial Gaussian and $\alpha_{0}$ sets the initial width. Note
that the initial choice of $q_{0}$ and $p_{0}$ also sets the values of $c_{1}$
and $c_{2}$ in Eq. (\ref{ct}), equal respectively to $q_{0}/\alpha_{0} $ and
$p_{0}\alpha_{0}/2\hbar m$.

From Eq. (\ref{psisol}) it appears that the exact solution of the
Schr\"{o}dinger equation with the initial condition (\ref{1}) has at
all times its maximal probability along the curve $q(t)$: this
defines the guiding trajectory. The evolution of the wavefunction
$\psi^{(q_{0},p_{0})}(x,t)$ depends on the properties of the guiding
trajectory (which as we have seen above, turns out to be the
solution obtained by solving the classical equations with the
Hamiltonian (\ref{ham})). The ensuing classical correspondence is in
line with Ehrenfest's theorem, though in a stronger form: in case in
which the initial wavefunction is a superposition of Gaussians
(\ref{1}) with different parameters $q_0$ and $p_0$, Ehrenfest's
theorem would then apply for each individual propagating wavepacket.

We will be interested in cases in which the time-dependent part of
the potential, $V(t)$ is periodic. The stability properties of the
general solutions of (\ref{er1}) -- Hill's equation -- are
well-known \cite{book nayfeh} by resorting to Floquet theory.\ For
the present purpose of this work, it will suffice to restrict the
discussion to the simplest non-trivial case, namely when the
time-dependence
takes the form%
\begin{equation}
V(t)=v-\kappa\cos\left(  2\omega t\right)
\end{equation}
Then Eq. (\ref{er1}) becomes a Mathieu equation and the guiding trajectory
$q(t)$ is given in terms of real even (``cosine'') and odd (``sine'') Mathieu
functions. It is well-known \cite{abramowitz} that for a given $\kappa$, the
solutions are bounded and periodic only for specific ``characteristic values''
of $v(\kappa)$.


\vspace{1cm}

\begin{thebibliography}{99}                                                                                               %
\bibitem {schulman}L. S. Schulman 1996, Techniques and Applications of Path
Integration (Wiley, New York).

\bibitem {bohm}P.\ R.\ Holland, 1995 \emph{\ The Quantum Theory of Motion} (Cambridge
Univ.\ Press, Cambridge).

\bibitem {QCbook}S.\ Wimberger 2014 \emph{Nonlinear Dynamics and Quantum Chaos}
(Springer, Heidelberg).

\bibitem {chaos}S. Freund, R. Ubert, E. Fl\"{o}thmann, K. Welge, D. M. Wang,
and J. B. Delos 2002, Phys. Rev. A 65, 053408; C. Bordas, F. L\'{e}pine, C.
Nicole, and M. J. J. Vrakking 2003 Phys. Rev. A 68, 012709. A. Matzkin, M.
Raoult and D. Gauyacq 2003 Phys. Rev. A 68 , 061401(R);
M. Lebental, N. Djellali, C. Arnaud, J.-S. Lauret, J. Zyss, R. Dubertrand, C.
Schmit, and E. Bogomolny 2007 Phys. Rev. A 76, 023830; J. D. Wright, J. M.
DiSciacca, J. M. Lambert, and T. J. Morgan 2010, Phys.Rev. A 81 , 063409 (2010); W.
Gao, H. F. Yang, H. Cheng, X. J. Liu, and H. P. Liu 2012 Phys. Rev. A 86, 012517.

\bibitem {dbbapp}Y. Nogami, F.M. Toyama, W. van Dijk 2000, Phys.\ Lett.\ A 270, 279;
 J. Acacio de Barros, G. Oliveira-Neto, and  T.B. Vale 2005 Phys. Lett A 336 324
 A.\ Matzkin 2005, Phys. Lett. A 345, 31; X.\ Oriols 2007 Phys. Rev. Lett.
98, 066803; X. Y. Lai, Qing-yu Cai and M. S. Zhan 2009 New J. Phys. 11,
113035; S.\ Mousavi 2010 J. Phys. A: Math. Theor. 43 035304;

\bibitem {wyatt}R.\ E.\ Wyatt 2005 \emph{Quantum Dynamics with
Trajectories} (Springer, New York).

\bibitem {mismatch}A. Matzkin and V. Nurock 2008, Studies in Hist. and Phil. of
Science B 39, 17; A. Matzkin 2009, Found. Phys. 39, 903.

\bibitem {AAV}Y. Aharonov, D. Z. Albert, and L. Vaidman, Phys. Rev. Lett. 60,
1351 (1988).

\bibitem {WMref1}A. G. Kofman, S. Ashhab and F. Nori, Phys. Rep. 520, 43
(2012).

\bibitem {WMref2}J. Dressel, M. Malik, F. M. Miatto, A. N. Jordan, and R. W.
Boyd 2014 Rev. Mod. Phys. 86, 307.

\bibitem {WMref3}A. Hosoya and Y.Shikano 2010 J. Phys. A 43 385307.

\bibitem {leavens}C.\ R.\ Leavens 2005, Found.\ Phys.\ 35, 469.

\bibitem {wiseman07}H. M. Wiseman 2007, New J. Phys. 9 165.

\bibitem {hiley}B. J. Hiley 2012, J. Phys.: Conf. Ser. 361 012014.

\bibitem {robust}F. L. Traversa, G. Albareda, M. Di Ventra, and X. Oriols 2013,
Phys. Rev. A 87 052124.

\bibitem {kocsis}S. Kocsis, B. Braverman, S. Ravets, M. J. Stevens, R. P.
Mirin, L. K. Shalm and A. M. Steinberg 2011, Science 332, 6034, 1170.

\bibitem {weak trajectories}A. Matzkin 2012, Phys. Rev. Lett. 109, 150407.


\bibitem {tsutsui}T. Mori and I. Tsutsui 2014, arXiv:1412.0916v1.


\bibitem {rmp}D. Leibfried, R. Blatt, C. Monroe, and D. Wineland 2003, Rev. Mod.
Phys. 75, 281

\bibitem {photon}V. V. Dodonov, M. A. Marchiolli, Ya. A. Korennoy, V. I.
Man'ko, and Y. A. Moukhin 1998, Phys. Rev. A 58, 4087

\bibitem {cosmo}S. P. Kim and S. W. Kim 1995, Phys. Rev. D 51 4254.

\bibitem {pan12}A. K. Pan and A. Matzkin 2012, Phys. Rev. A 85, 022122

\bibitem{joza} R. Jozsa 2007 Phys. Rev. A 76, 044103

\bibitem {supplem}See the Supplemental Material of Ref. \cite{weak
trajectories}.

\bibitem {lawande}S.\ V.\ Lawande and A.\ K.\ Dhara 1983, Phys.\ Lett.\ A 99 353.

\bibitem {dbb}L.\ de Broglie 1928, in \emph{Electrons et Photons}
(Gauthier-Villars, Paris), p. 105; D.\ Bohm 1952, Phys. Rev. 85, 166.

\bibitem{hileycallaghan} B. J. Hiley and R. E. Callaghan 2012, Found. Phys. 42, 192.

\bibitem {hiley2006}B. J. Hiley and R. E. Callaghan 2006, Phys. Scr. 74, 336.

\bibitem {alg}C.\ F.\ Lo 1993, Phys. Rev. A 47, 115.

\bibitem {LR}H. R. Lewis and W. B. Riesenfeld 1969 J. Math. Phys. 10, 1458.

\bibitem {moshinskyPRA}D. Schuch and M. Moshinsky 2006, Phys. Rev. A 73, 062111.

\bibitem {ermakov}J.\ R.\ Ray and J.\ L.\ Reid, J. Math. Phys. 22, 91 (1981);
A. Matzkin 2001, J. Phys. A 34, 7833;
F Haas 2002 J. Phys. A 35 2925;
M Fernández Guasti and H Moya-Cessa 2003 J. Phys. A 36 2069.

\bibitem {book nayfeh}A. H. Nayfeh and D. T. Mook 1995, \emph{Nonlinear
oscillations} (Wiley, New York).

\bibitem {abramowitz}M. Abramowitz and I. A. Stegun (Eds) 1964, \emph{Handbook of
Mathematical Functions} (National Bureau of Standards, Washigton DC), Chap.\ 20.
\end{thebibliography}
\end{document}